\documentclass[11pt]{article}

\usepackage[final]{acl}
\usepackage{times}
\usepackage{latexsym}
\usepackage[T1]{fontenc}
\usepackage[utf8]{inputenc}
\usepackage{microtype}
\usepackage{inconsolata}

\usepackage{graphicx}
\usepackage{float}

\usepackage{tabularx}
\usepackage{xcolor}
\usepackage{soul}
\usepackage{booktabs}
\usepackage{multirow}

\usepackage[linesnumbered,ruled,vlined]{algorithm2e}
\SetAlCapFnt{\small}
\SetAlCapNameFnt{\small}

\newcommand{\subdef}[1]{\item \textnormal{#1:}}
\usepackage{amsthm}
\usepackage{amssymb}
\usepackage{amsmath}
\usepackage{bbold}
\usepackage{dsfont}
\newtheoremstyle{customdefinition}{5pt}{5pt}{}{}{\bfseries}{.}{1em}
    {\thmname{#1}\thmnumber{ #2} \thmnote{ \textnormal{\textbf{(#3)}}}}
\theoremstyle{customdefinition}
\newtheorem{definition}{Definition}[section]
\newtheorem{theorem}{Theorem}[section]

\newtheorem{assumption}{Assumption}[section]

\usepackage{cleveref}
\AtBeginDocument{
    \crefformat{appendix}{Appendix~#2#1#3}
    \crefformat{subappendix}{Appendix~#2#1#3}
    \crefformat{section}{Section~#2#1#3}
    \crefformat{subsection}{Subsection~#2#1#3}
    \crefformat{subsubsection}{Subsubsection~#2#1#3}
    \crefformat{figure}{Figure~#2#1#3}
    \crefformat{subfigure}{Figure~#2#1#3}
    \crefformat{table}{Table~#2#1#3}
    \crefformat{equation}{Equation~(#2#1#3)}
    \crefformat{definition}{Definition~#2#1#3}
    \crefformat{theorem}{Theorem~#2#1#3}
    \crefformat{lemma}{Lemma~#2#1#3}
    \crefformat{corollary}{Corollary~#2#1#3}
    \crefformat{assumption}{Assumption~#2#1#3}
    \crefformat{proposition}{Proposition~#2#1#3}
    \crefformat{algorithm}{Algorithm~#2#1#3}
}

\usepackage{comment}
\renewenvironment{comment}{}{}

\usepackage[caption=false,font=footnotesize]{subfig}
\newcommand{\model}{\mathsf{M}}
\newcommand{\testSuite}{\mathsf{T}}
\newcommand{\watermark}{\mathsf{Watermark}}
\newcommand{\detect}{\mathsf{Detect}}
\newcommand{\secretKey}{\mathsf{k}}
\newcommand{\wmScheme}{\mathsf{\Pi}}
\newcommand{\adversary}{\mathsf{Att}}
\newcommand{\eqRule}{\mathsf{R}}
\newcommand{\graph}{\mathsf{G}}
\newcommand{\trans}{\mathsf{Transform}}
\newcommand{\step}{\mathsf{RandomStep}}
\newcommand{\walk}{\mathsf{RandomWalk}}
\newcommand{\stepNum}{t}
\newcommand{\func}{\mathcal{F}}
\newcommand{\prompt}{x}
\newcommand{\code}{c}
\newcommand{\action}{a}
\newcommand{\vertex}{v}
\newcommand{\edge}{e}
\newcommand{\weight}{w}
\newcommand{\ruleSet}{\mathit{\Gamma}}
\newcommand{\ergRuleSet}{\ruleSet_{\mathsf{ERG}}}
\newcommand{\actionSet}{\mathit{\Lambda}}
\newcommand{\partitionSimple}{\mathcal{P}}
\newcommand{\partition}{\partitionSimple^{\ergRuleSet}}
\newcommand{\vertexSet}{\mathit{V}}
\newcommand{\edgeSet}{\mathit{E}}
\newcommand{\weightSet}{\mathit{W}}
\newcommand{\transMat}{\vec{\mathit{P}}}
\newcommand{\statDist}{\pi}
\newcommand{\distVec}{\mathit{p}}
\newcommand{\expect}{\mathds{E}}
\newcommand{\indicator}{\mathbb{1}}
\newcommand{\nature}{\mathbb{N}}
\newcommand{\real}{\mathbb{R}}
\newcommand{\sizeof}[1]{|#1|}
\newcommand{\tvDist}[2]{\|#1 - #2\|_{\text{TV}}}
\newcommand{\bern}{\text{Bernoulli}}

\newcommand{\normal}{\mathcal{N}}
\newcommand{\unknownDist}{\mathcal{D}}
\newcommand{\spaceDist}{\mathit{r}}
\newcommand{\promptSpace}{\mathcal{X}}
\newcommand{\codeSpace}{\mathcal{C}}
\newcommand{\modelSpace}{\mathcal{M}}
\newcommand{\keySpace}{\mathcal{K}}
\newcommand{\actionSpace}{\mathcal{A}}
\newcommand{\eqSpace}{\Phi}
\newcommand{\ruleSpace}{\mathcal{R}}

\newcommand{\funcErgEqSpace}{\eqSpace^{\ergRuleSet}}
\newcommand{\hqSpace}{\eqSpace^{\func}}
\newcommand{\fnr}{\epsilon_{\mathsf{neg}}}
\newcommand{\fpr}{\epsilon_{\mathsf{pos}}}
\newcommand{\aRate}{\epsilon}
\newcommand{\curly}[1]{\left\{#1\right\}}
\begin{document}

\title{Disappearing Ink: Obfuscation Breaks N-gram Code Watermarks in Theory and Practice}

\author{
    Gehao Zhang \quad
    Mingzhe Li \quad
    Eugene Bagdasarian \quad
    Shiqing Ma \quad
    Juan Zhai \\
    Manning College of Information and Computer Sciences \\
    University of Massachusetts Amherst \\
    Amherst, MA, USA \\
    \texttt{\{gehaozhang, mingzhel, eugene, shiqingma, juanzhai\}@umass.edu}
}

\maketitle


\begin{abstract}

Large language models (LLMs) are increasingly used for code generation, making reliable identification of machine-generated code important for attribution, tracking, and misuse detection. Existing code watermarking methods are dominated by N-gram-based schemes, yet their robustness has mostly been evaluated only against simple edits or optimizations. We argue that this significantly overstates security, because software engineering \emph{already} provides stronger semantics-preserving transformations in the form of code obfuscation.

We study N-gram-based code watermarking under obfuscation. We formally model semantics-preserving transformations as a Markov random walk and prove that, under an intuitive and experimentally supported assumption called \textit{distribution consistency}, obfuscation can nullify the robustness of N-gram-based watermarks. If the original detector has a false positive rate $\fpr$, then after obfuscation, its failure rate on watermarked code approaches $1 - \fpr$.

We validate this theory on three state-of-the-art watermarking schemes, two LLMs, two programming languages, four benchmarks, and four obfuscators. Across all settings, detectors collapse to near-random performance on obfuscated code (AUROC tightly around $0.5$), and for each language, at least one attack leaves all post-obfuscation AUROC scores below $0.6$. These results jointly show that current N-gram-based code watermarks are not robust to realistic obfuscation attacks and motivate more semantics-aware alternatives.

\end{abstract}

\section{Introduction}

\begin{figure}[t]
    \centering
    \hspace{-20pt}
    \includegraphics[width=0.85\linewidth]{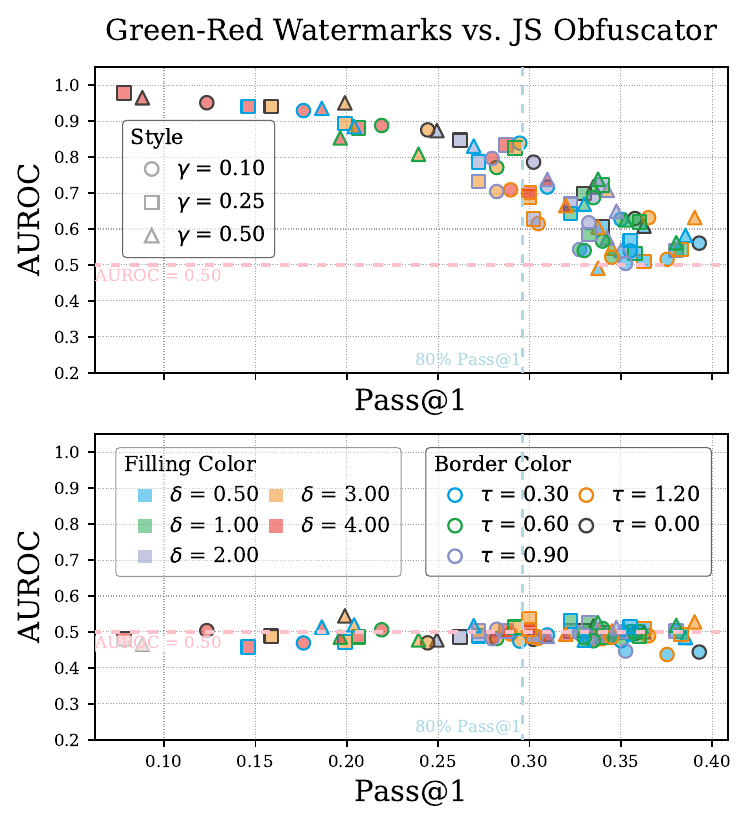}
    \caption{
    Green-Red watermark performance before (top) and after (bottom) our JS Obfuscator attack on DeepSeek-Coder. 
    The y-axis shows detection AUROC, and the x-axis is the Pass@1 score, representing code generation ability.
    Different point styles denote different values of hyperparameters.
    }
    \label{fig:green_red_intro}
\end{figure}
As LLM-generated content becomes pervasive, keeping it distinguishable from human-created content is becoming a practical requirement. Recent policy and regulatory efforts around AI labeling and watermarking reflect exactly this demand \cite{luo2025china, heitzenrater2024watermarking, catacora2024california}. In software engineering, where LLMs are rapidly adopted for code generation \cite{chen2021evaluating, li2022competition, wu2024devin, islam2024mapcoder}, watermarking is a natural candidate for providing such distinguishability. Representative schemes such as SynthID \cite{dathathri2024scalable}, WLLM \cite{kirchenbauer2023watermark}, and SWEET \cite{lee2023wrote} aim to provide that distinguishability while balancing detectability and code quality, and some have already seen industrial deployment \cite{dathathri2024scalable}. Despite their implementation differences, these methods largely follow the same pattern: they hash the previous $(N-1)$ tokens into a pseudo-random seed that affects the next-token choice, and the detector later relies on intact N-grams to recover that seed. This pattern is widely followed by modern watermarking methods, which we refer to as \textit{N-gram-based} watermarking \cite{dathathri2024scalable, kirchenbauer2023watermark, lee2023wrote, kirchenbauer2023reliability, liu2023unforgeable, wu2023resilient, zhao2024permute, fu2024gumbelsoft}.

\begin{figure}[t]
    \centering
    \includegraphics[width=0.8\columnwidth]{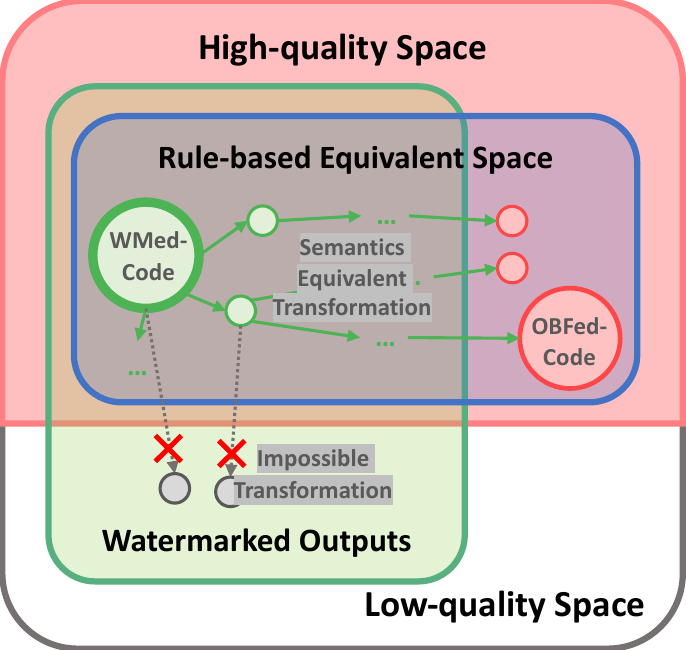}
    \caption{
    An outline of our attack schema.
    Our attack randomly modifies code using a set of semantics-preserving transformation rules, i.e., code obfuscation.
    We prove that starting from a high-quality watermarked code sample, the obfuscated code converges to a distribution within a \emph{subset} of the high-quality space (\cref{tho:inside_ind}).
    Furthermore, the resulting partition is largely independent of N-gram features, leading to chance-level detection performance (\cref{tho:main}; \cref{fig:green_red_main}).
    }
    \label{fig:theroy_model}
\end{figure}

\begin{figure*}[t]
    \setlength{\fboxsep}{1pt}
    \definecolor{hlgreen}{HTML}{57a95e}
    \definecolor{textgreen}{HTML}{186918}
    \definecolor{presblue}{HTML}{2F5597}
    \newcommand{\hlt}[1]{
        \colorbox{hlgreen}{\textcolor{textgreen}{\texttt{#1}}}
    }
    \newcommand{\obfn}[1]{
        \textcolor{red}{#1}
    }
    \newcommand{\pres}[1]{
        \underline{\textcolor{presblue}{\texttt{#1}}}
    }
    \centering
    \includegraphics[width=1\textwidth]{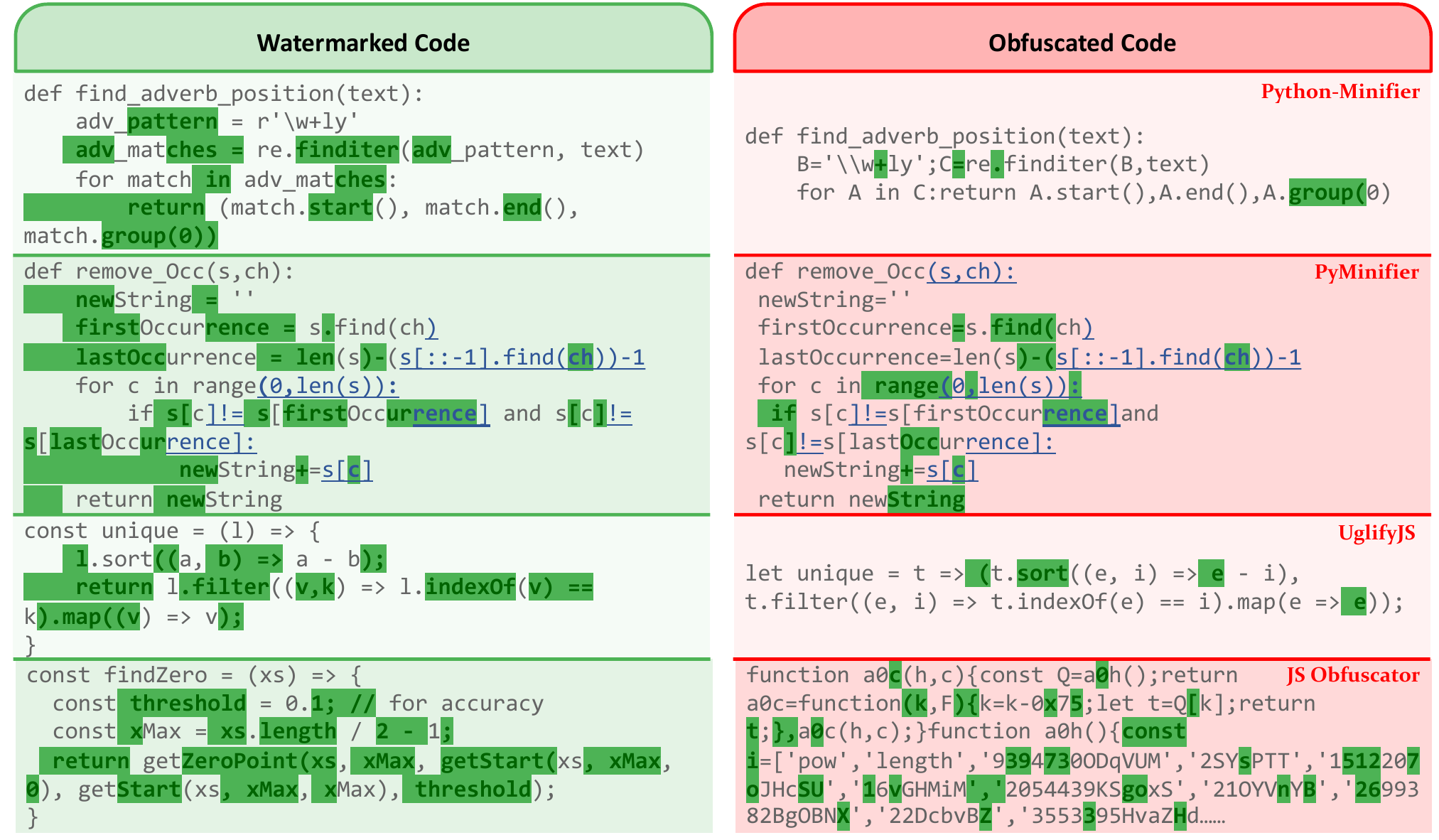}
    \caption{
    Our attack on watermarked code \cite{kirchenbauer2023watermark}.
    The left panel contains code samples generated by the watermarked model. The watermarking scheme adds hidden strategies to generate more \hlt{green tokens} that the detector counts to judge whether the text was watermarked.
    The right panel includes the corresponding code after obfuscation. The fraction of \hlt{green tokens} decreases.
    The \obfn{red} labels show the names of the obfuscators, which are designed to preserve the semantics of these code pairs.
    A token in \pres{blue with an underline} ends a 5-gram that is preserved across obfuscation, and therefore keeps the green/red assignment it received at generation time; a preserved token is not necessarily a \hlt{green token}.
    }
    \label{fig:wm_examples}
\end{figure*}

\begin{figure}[b]
    \newcommand{\preserve}[1]{
        \textcolor{blue}{#1}
    }
    \centering
    \includegraphics[width=1\linewidth, trim={0 0mm 0 0}, clip]{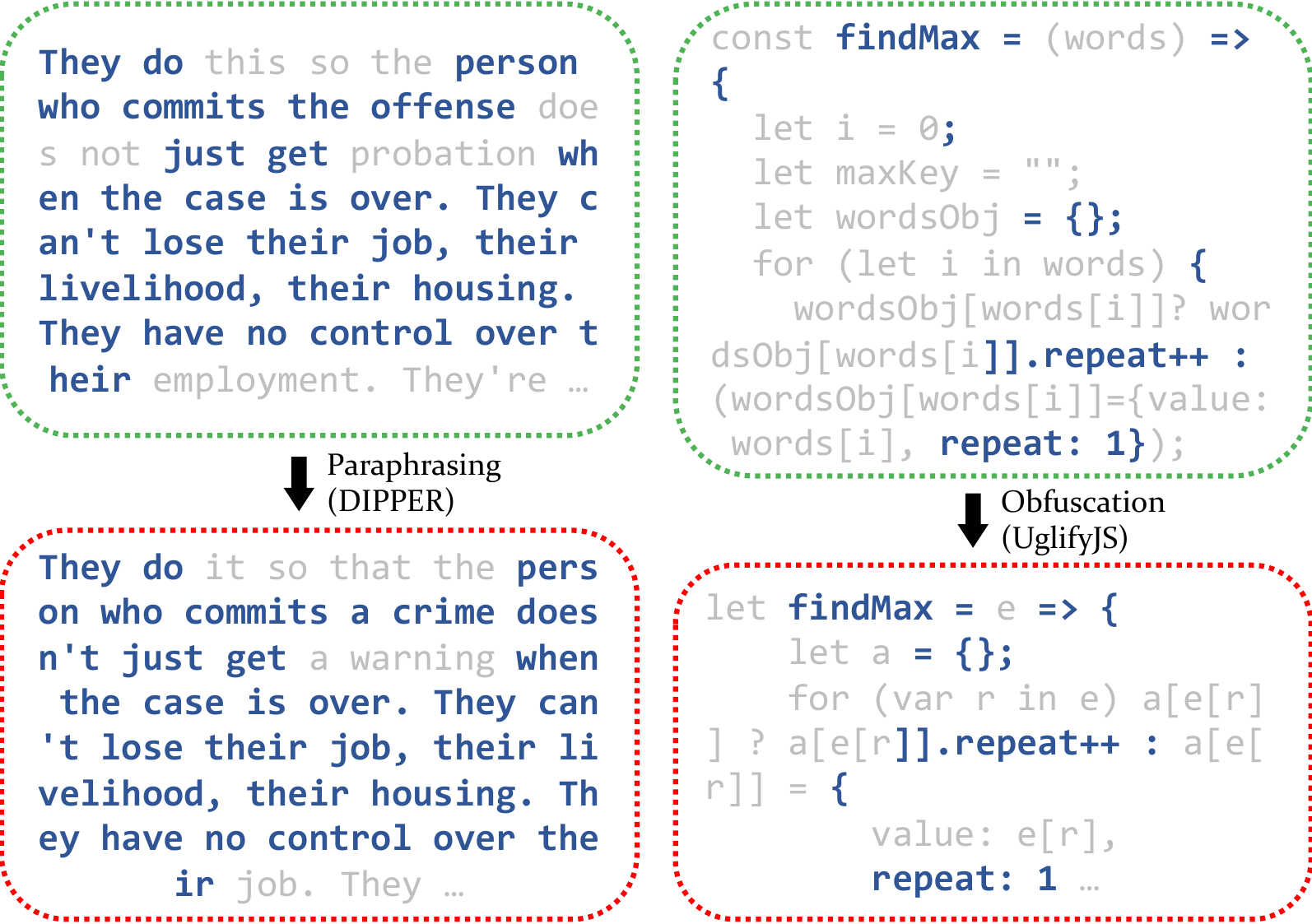}
    \caption{
    Comparison between natural language paraphrasing and code obfuscation. The \preserve{blue} color indicates the 2-grams preserved after the modification.
    }
    \label{fig:nl_comp}
\end{figure}

For watermarking, detectability without robustness is of limited value. A scheme is useful only if its signal survives realistic post-perturbation or modification.
In natural language, paraphrasing attacks \cite{krishna2023paraphrasing, sadasivan2023can} have already been used to disrupt watermarks in natural language. 
For code, the attack surface is even stronger: semantics-preserving rewrites such as renaming, dead-code insertion, and control-flow changes are abundant, and mature code obfuscators already automate them \cite{banescu2018tutorial, behera2015different, shah2018code, peng2019adaptive, you2010malware}. As illustrated in \autoref{fig:nl_comp}, code obfuscation can preserve semantics more reliably than natural-language paraphrasing while disturbing N-grams more aggressively. Yet prior code watermark evaluations mostly rely on weaker edits, optimizations, or refactorings rather than realistic obfuscation attacks \cite{lee2023wrote, li2024resilient, ning2024mcgmark, guan2024codeip}.
This gap motivates our attack model. While prior impossibility results for general language models exist \cite{zhang2023watermarks}, their assumptions are not well suited to code generation (\autoref{sec:imp_of_imp}). We instead give an implementable, ergodic attack based on semantics-preserving rule-based transformations. The model, outlined in \autoref{fig:theroy_model}, treats obfuscation as a Markov random walk over equivalent programs.
\cref{fig:wm_examples} exhibits real watermarked code before and after such an attack.

Our main theoretical result is that, under the intuitive and empirically supported assumption of \textit{distribution consistency}, obfuscation nullifies the robustness of every N-gram-based watermarking scheme (\autoref{tho:main}). If the original detector has a false positive rate $\fpr$, then its failure rate on obfuscated watermarked code rises to $1 - \fpr$ (\autoref{tho:main}, \autoref{tho:lower_bound}), meaning the detector loses discriminative power.

We instantiate the attack with off-the-shelf obfuscators and evaluate three watermarking schemes across two LLMs, two programming languages, four benchmarks, and four obfuscators. In \autoref{fig:green_red_intro}, after attack, all hyperparameter settings cluster tightly around AUROC $0.5$, i.e., near chance-level detection, while preserving code semantics; the same post-obfuscation collapse holds across all settings, with no AUROC above $0.6$.
Relying on cheap, CPU-only, off-the-shelf tools is a deliberate part of the threat model: the easier an attack is to mount, the more serious the vulnerability it exposes.
Meanwhile, our experiment shows that \textit{distribution consistency} holds in 96.11\% of the $1{,}620$ tested equivalent spaces (Section \ref{sec:dist_test_ext}).

Source code and dataset are available at~\url{https://github.com/zhang-ge-hao/CodeWM}.

\section{Related Work}


\paragraph{Large Language Model Watermarking.}
Modern LLM watermarking embeds signals directly into generation \cite{zhao2024sok}. Representative families include Green-Red watermarking \cite{kirchenbauer2023watermark, zhao2023provable, kirchenbauer2023reliability, lee2023wrote, liu2023unforgeable, liu2024adaptive, huo2024token, zhou2024bileve}, Gumbel watermarking \cite{aaronson2023watermarking, kuditipudi2023robust, hu2023unbiased, wu2023resilient, zhao2024permute, fu2024gumbelsoft}, and SynthID \cite{dathathri2024scalable}.
Despite these design differences, many strong methods remain N-gram-based \cite{dathathri2024scalable, kirchenbauer2023watermark, lee2023wrote, kirchenbauer2023reliability, liu2023unforgeable, wu2023resilient, zhao2024permute, fu2024gumbelsoft}.

\paragraph{Code Obfuscation.}
Code obfuscation transforms a program into a functionally equivalent but less readable version. Existing obfuscators aggressively rewrite code while preserving behavior, therefore providing a realistic threat model for code watermarking \cite{wong2006hunting, konstantinou2008metamorphic, you2010malware}.

\paragraph{Paraphrasing Attack.}
Paraphrasing attacks construct semantically similar adversarial text to break brittle token patterns in NLP systems \cite{merkhofer2022practical, ebrahimi2018adversarial, cheng2019robust, wallace2019universal, cheng2020seq2sick, zhao2017generating, zhang2021crafting}. For watermarking, DIPPER and recursive paraphrasing both expose fragility in current detectors \cite{krishna2023paraphrasing, kirchenbauer2023watermark, mitchell2023detectgpt, gptzero2024ai, krishna2022rankgen, sadasivan2023can}. We study the analogous robustness question for code, where semantics-preserving rewrites are more structured and often stronger.

\paragraph{Watermarking Impossibility.}
Prior work gives impossibility arguments for watermark robustness in general language-generation settings \cite{zhang2023watermarks}. However, their assumptions do not directly fit code generation, where attacks must preserve program semantics (\autoref{sec:imp_of_imp}). We extend this line with an implementable, semantics-preserving attack model for code.
\section{Theoretical Model and Impossibility Result}
\label{sec:theory}

This section gives the main theoretical argument behind our attack.
The full formalization and detailed proofs are in Appendices~\ref{apx:sec:theory_details}--\ref{apx:sec:impossibility}.
At a high level, we model semantics-preserving code obfuscation as a random walk over a graph: each vertex is a high-quality program, and each edge is a valid equivalent transformation.
For any fixed reachable equivalent space, standard Markov-chain convergence implies that, after sufficient mixing, the final program is independent of the particular starting program within that space.
To reason across different reachable spaces, we further introduce a distribution-consistency assumption: for N-gram-based watermarking, realistic rule-based transformations induce reachable spaces that should be independent of the detector's watermark signal.
We support this assumption with discussion and experiments.
Let $\fpr$ denote the false-positive rate of the original watermarking scheme.
Assuming distribution consistency, the detector identifies obfuscated watermarked code with probability $\fpr$, and thus misses it with probability approximately $1-\fpr$.

\subsection{Threat Model}
\label{sec:threat_model}

Let $\promptSpace$ be the prompt space and $\codeSpace$ be the code space. A randomized code model $\model:\promptSpace\to\codeSpace$ generates code for a prompt $\prompt$. A secret-key watermarking scheme $\wmScheme=(\watermark,\detect)$ first produces a key $\secretKey\in\keySpace$ and a watermarked model $\model_{\secretKey}$, and then uses a deterministic detector $\detect_{\secretKey}(\prompt,\code)\in\curly{0,1}$, where $1$ means that the code is judged as watermarked.
For a fixed prompt $\prompt$, let $\hqSpace\subseteq\codeSpace$ denote the high-quality code space: programs that satisfy the prompt-required behavior, e.g., by passing the intended tests. 

An attacker receives a watermarked high-quality program $\code\in\hqSpace$ and outputs another program $\adversary(\prompt,\code)\in\hqSpace$.  In other words, the attack is only allowed to modify the program while preserving its functional behavior.
The attack goal is to make a watermarked program evade detection:
\[
\begin{aligned}
    \Pr[\detect_{\secretKey}(\prompt,\code')=0
    \mid{}& \code\leftarrow \model_{\secretKey}(\prompt),\\
    &\code'=\adversary(\prompt,\code)].
\end{aligned}
\]
A robust watermark should keep this probability low even after attacks.

\subsection{Random Walk over Equivalent Programs}
\label{sec:eq_trans}
\label{sec:chain_modeling}

A code obfuscator can be viewed as a collection of local semantics-preserving transformation rules: renaming variables, adding or removing comments, inserting or deleting dead code, rewriting expressions, or replacing one control-flow form by another equivalent form.  We abstract these operations as a rule set $\ruleSet$.
Given a program $\code$, $\ruleSet$ identifies a set of currently applicable semantics-preserving transformations.

\begin{definition}[Random Transformation and Rule-based Equivalent Space]
\label{def:random_transformation}
\label{def:rule_based_eq_space}
Given a rule set $\ruleSet$, $\step(\code,\ruleSet)$ randomly chooses one applicable transformation action from $\ruleSet$ and applies it to $\code$.
The algorithm $\walk(\code,\ruleSet,\stepNum)$ repeats $\step$ for $\stepNum$ steps.
We write $\code \leadsto_{\ruleSet} \code'$ if $\code'$ can be reached from $\code$ with non-zero probability after finitely many steps.

For an initial high-quality program $\code\in\hqSpace$, the rule-based equivalent space is
\[
    \eqSpace^{\ruleSet}_{\code}
    =
    \{\,\code' \in \hqSpace
    \mid
    \code \leadsto_{\ruleSet} \code'\,\}.
\]
That is, $\eqSpace^{\ruleSet}_{\code}$ contains all high-quality programs reachable from $\code$ by finite steps in $\ruleSet$.
See \cref{apx:def:random_transformation} for the definition in detail.
\end{definition}

The equivalent space naturally defines a directed weighted graph $\graph^{\ruleSet}_{\code}=(\vertexSet,\edgeSet,\weightSet)$: every reachable program is a vertex; there is an edge from $\code_i$ to $\code_j$ if $\Pr[\code_j=\step(\code_i,\ruleSet)]>0$; and the edge weight is exactly this one-step probability.  Therefore, repeatedly applying $\step$ is a Markov chain over $\eqSpace^{\ruleSet}_{\code}$.

\begin{definition}[Ergodic Transformation Rule Set]
\label{def:erg_rule_set}
A rule set $\ergRuleSet$ is ergodic if it contains a do-nothing transformation and every transformation can be reversed with non-zero probability.  Equivalently, inside any reachable equivalent space, the induced transformation graph is connected through reversible paths and has self-loops.
See \cref{apx:def:erg_rule_set} for the definition in detail.
\end{definition}

This condition is not merely theoretical.  Many common obfuscation operations have this form: renamed variables can be renamed back; comment insertion and deletion reverse each other; and dead-code insertion and deletion reverse each other.  The do-nothing rule is included only to avoid periodicity in the random walk.

\begin{theorem}[Inside-space Independence]
\label{tho:inside_ind}
Let $\distVec(\code,\stepNum)$ be the distribution over vertices after $\stepNum$ steps starting from $\code$.
For any $\code\in\hqSpace$ and any ergodic rule set $\ergRuleSet$, the Markov chain induced by $\walk(\cdot,\ergRuleSet,\stepNum)$ over $\eqSpace^{\ergRuleSet}_{\code}$ has a unique stationary distribution $\statDist$.  Moreover,
\[
    \lim_{\stepNum\to\infty}
    \distVec(\code',\stepNum)=\statDist,
    \quad
    \code'\in \eqSpace^{\ergRuleSet}_{\code} .
\]
In words, after sufficiently many transformations, the distribution of the final program depends only on the reachable equivalent space, not on the exact starting program inside that space.
\end{theorem}

\noindent\textit{Proof sketch.}
Because every program in $\eqSpace^{\ergRuleSet}_{\code}$ is reachable from the initial program by definition, and because every rule can be reversed with non-zero probability, any two programs in the same equivalent space can reach each other through the initial program.  Thus the chain is irreducible.  The do-nothing transformation gives every vertex a self-loop, so the chain is aperiodic.  A finite irreducible and aperiodic Markov chain has a unique stationary distribution and converges to it from every starting point.
See \cref{apx:tho:inside_ind} for the full proof.

The finiteness used here concerns only the reachable component $\eqSpace^{\ergRuleSet}_{\code}$, not the set of all programs syntactically equivalent to $\code$.
Even when unrestricted renaming or comment insertion would generate unboundedly many equivalent programs, the rule set can be designed so that every walk stays within a bounded state space.
Our reference obfuscator (\cref{sec:rw_obf}) realizes this by confining all transformations of a seed program to a fixed $2{,}000$-character length interval, which keeps the chain finite while preserving irreducibility and aperiodicity.

An immediate consequence is that $\ergRuleSet$ partitions the high-quality space into disjoint rule-based equivalent spaces: two programs either belong to the same reachable component, or their reachable components do not overlap.  We write this partition as $\partition=\curly{\eqSpace_1,\ldots,\eqSpace_m}$.

\begin{theorem}[Rule-based Partition]
\label{tho:partition}
For a fixed ergodic rule set $\ergRuleSet$, the family of spaces $\eqSpace^{\ergRuleSet}_{\code}$ forms a partition of $\hqSpace$: if two spaces overlap, they are the same space; or they are disjoint.
\end{theorem}

\noindent\textit{Proof sketch.}
If two spaces share a program, reversibility makes each space reachable from the other through it; hence the spaces are identical.
A full proof is in \cref{apx:tho:partition}.

\subsection{Distribution Consistency for N-gram Watermarks}
\label{sec:impossibility}

The previous theorem says what happens \emph{inside} each rule-based equivalent space: the random walk forgets the starting program.
In this section, we extend the independence argument from a single rule-based equivalent space $\eqSpace^{\ergRuleSet}_{\code}$ to the entire high-quality code space $\hqSpace$.

\begin{definition}[Distribution Consistency]
\newcommand{\anyDist}{\mathcal{Q}}
\label{def:consistency}
Given a rule-based partition $\partition$, a detector
$\detect_{\secretKey}$ satisfies distribution consistency if conditioning on
the equivalent space does not change the distribution of the detector's
output.  Formally, for any code distribution $\anyDist$ independent of the
watermarking scheme, if
$\code\sim\anyDist(\hqSpace)$ and
$\detect_{\secretKey}(\prompt,\code)\sim\unknownDist$, then for every
$\eqSpace_j\in\partition$,
\[
    \code\sim\anyDist(\eqSpace_j)
    \quad\Longrightarrow\quad
    \detect_{\secretKey}(\prompt,\code)\sim\unknownDist .
\]
That is, the distribution of $\detect_{\secretKey}$ within each equivalent space $\eqSpace \in \partition$ remains the same as the distribution observed over the entire code space.
\end{definition}

Note that distribution consistency is a necessary and sufficient condition for the independence between $\ergRuleSet$  and $\detect$.

\begin{assumption}[Impossibility Assumption]
\label{asm:main}
For N-gram-based code watermarking, there exist implementable semantics-preserving transformation rules whose induced rule-based partition satisfies distribution consistency.
\end{assumption}

This is the only extra assumption in our main argument.  It should not be read as saying that \emph{every} rule set breaks \emph{every} detector.  A weak rule set, such as only doing nothing, clearly preserves the watermark signal.  Instead, the assumption says that realistic code transformations can perturb many N-gram-level features while preserving functional behavior.  Such transformations can change variable names, comments, constants, expressions, and local syntax, whereas the remaining stable information is mostly semantic.  Since N-gram-based watermarks are designed around local token patterns rather than program semantics, the N-gram signal is expected to become approximately independent of the rule-based equivalent-space partition.  We empirically test this assumption in \cref{sec:dist_test_ext}.

The boundary between what is proven and what is assumed can be stated precisely.
Within a single equivalent space, the behavior is proven: \cref{tho:inside_ind} shows that once the space containing the starting program is fixed, the walk converges to a stationary distribution $\statDist_j$ determined by the rule set and the graph topology alone, and hence independent of the watermarking scheme.
Across spaces, the behavior is assumed: \cref{def:consistency} applies to a scheme-independent distribution whose conditional distribution on each space $\eqSpace_j$ is exactly $\statDist_j$, and distribution consistency is the single extra hypothesis that the detector's output distribution within each space matches its distribution over the whole high-quality space.
The proofs of \cref{tho:main} and its appendix counterpart (\cref{apx:tho:main}) invoke the assumption at exactly one step, when the per-space detection rate is replaced by $\fpr$.

\subsection{Robustness Impossibility}

We now combine the two pieces.  First, the random walk reaches a stationary distribution inside the equivalent space of the original watermarked program.  Second, under distribution consistency, sampling from that equivalent space makes the N-gram detector fire with probability $\fpr$, the same as on ordinary non-watermarked code.  Therefore, the transformed watermarked program evades detection with probability $1-\fpr$.

\begin{theorem}[Robustness Impossibility for N-gram Watermarking]
\label{tho:main}
Assume \cref{asm:main}.
For any N-gram-based watermarking scheme with false-positive rate $\fpr$, there exists an implementable ergodic transformation rule set $\ergRuleSet$ such that, for sufficiently large $\stepNum$, the attacker
\[
    \adversary(\prompt,\code)
    =
    \walk(\code,\ergRuleSet,\stepNum)
\]
makes the post-obfuscation false-negative rate approach $1-\fpr$:
\[
\begin{aligned}
    &\Pr[\detect_{\secretKey}(
    \prompt,\code')=0
    \mid \code\leftarrow \model_{\secretKey}(\prompt),\\
    &\hspace{3.2em}
    \code'=\adversary(\prompt,\code)]
    \approx 1-\fpr .
\end{aligned}
\]
\end{theorem}

\noindent\textit{Proof sketch.}
Consider any watermarked program $\code$ and the equivalent space $\eqSpace_j$ that contains it.  By \cref{tho:inside_ind}, after enough random-walk steps, the output distribution is the stationary distribution of $\eqSpace_j$, independent of the particular starting program.  By distribution consistency, a program sampled from $\eqSpace_j$ is detected as watermarked with probability $\fpr$.  Hence it is missed with probability $1-\fpr$.  Averaging over all possible starting watermarked programs, and therefore over all equivalent spaces they may fall into, keeps the same value because the weights over spaces sum to $1$:
\[
\begin{aligned}
&\Pr[\detect_{\secretKey}(\prompt,\code')=0
  \mid \code\leftarrow \model_{\secretKey}(\prompt),
  \code'=\adversary(\prompt,\code)] \\
&\quad
=1-\sum_{\eqSpace_j\in\partition}p_jq_j
=1-\fpr\sum_{\eqSpace_j\in\partition}p_j \\
&\quad
=1-\fpr.
\end{aligned}
\]
By distribution consistency, $q_j=\fpr$ for all $j$.
Here
$p_j=\Pr[\code\in\eqSpace_j\mid \code\leftarrow\model_{\secretKey}(\prompt)]$
and
$q_j=\Pr[\detect_{\secretKey}(\prompt,\code')=1\mid \code'\in\eqSpace_j]$.

The theorem shows that, under distribution consistency, sufficient obfuscation reduces post-attack detection to the detector's false-positive behavior.
When $\fpr$ is small, the post-attack false-negative rate approaches one; detectability disappears.
A full proof is in \cref{apx:tho:main}.

\begin{theorem}[Finite-step Lower Bound]
\label{tho:lower_bound}
Let $\stepNum_{i}(\aRate)$ be the mixing time satisfying
\[
    \tvDist{\distVec(\code_i,\stepNum_{i}(\aRate))}
    {\statDist}\leq\aRate .
\]
Under \cref{asm:main}, for any $\stepNum\geq\stepNum_{i}(\aRate)$, the same random-walk attacker achieves at least $1-\fpr-\aRate$ post-attack false negative rate.
\end{theorem}

\noindent\textit{Proof sketch.}
Before exact convergence, the random walk is within total variation distance $\aRate$ of the stationary distribution.  This can change the detector's positive probability by at most $\aRate$.  Since the stationary positive probability is $\fpr$ under distribution consistency, the finite-step positive probability is at most $\fpr+\aRate$, and the miss probability is at least $1-\fpr-\aRate$.
A full proof is in \cref{apx:tho:lower_bound}.

Note that this post-attack false negative rate, 
$1 - \aRate - \fpr$, is still too high to be acceptable.
Under the conditions used in the mixing-time calculation in 
\cref{tho:mixing_cal}, $\stepNum_i(\aRate)$ grows linearly with 
$\ln \aRate^{-1}$.
Therefore, decreasing $\aRate$ improves the lower bound on the 
post-attack false negative rate while incurring only a logarithmic 
overhead in the required number of random-walk steps.

\subsection{Summary}

In summary, under a mild and experimentally supported assumption of \textit{distribution consistency}, obfuscation can effectively defeat N-gram-based watermarking schemes. In particular, we prove that with our assumption satisfied, the attack will make a fraction of approximately $1 - \fpr$ of watermarked code evade detection, which means that the detector entirely loses its ability to distinguish watermarked code from benign code.

While the obfuscation attack fundamentally challenges the robustness of existing watermarking schemes, we also discuss a possible defense path, i.e., making code watermarking more semantics-aware to break distribution consistency, with a proof-of-concept design in pseudocode (see Appendix \ref{sec:sem_aware}).
In addition, we present the space-/time-complexity calculation and the mixing-time estimation for our attack (see Appendix \ref{sec:cost_analysis}).
\section{Experimental results}
\label{sec:experimental_results}

\begin{figure*}[t]
    \centering
    \includegraphics[width=\textwidth, trim={0 2.7mm 0 0}, clip]{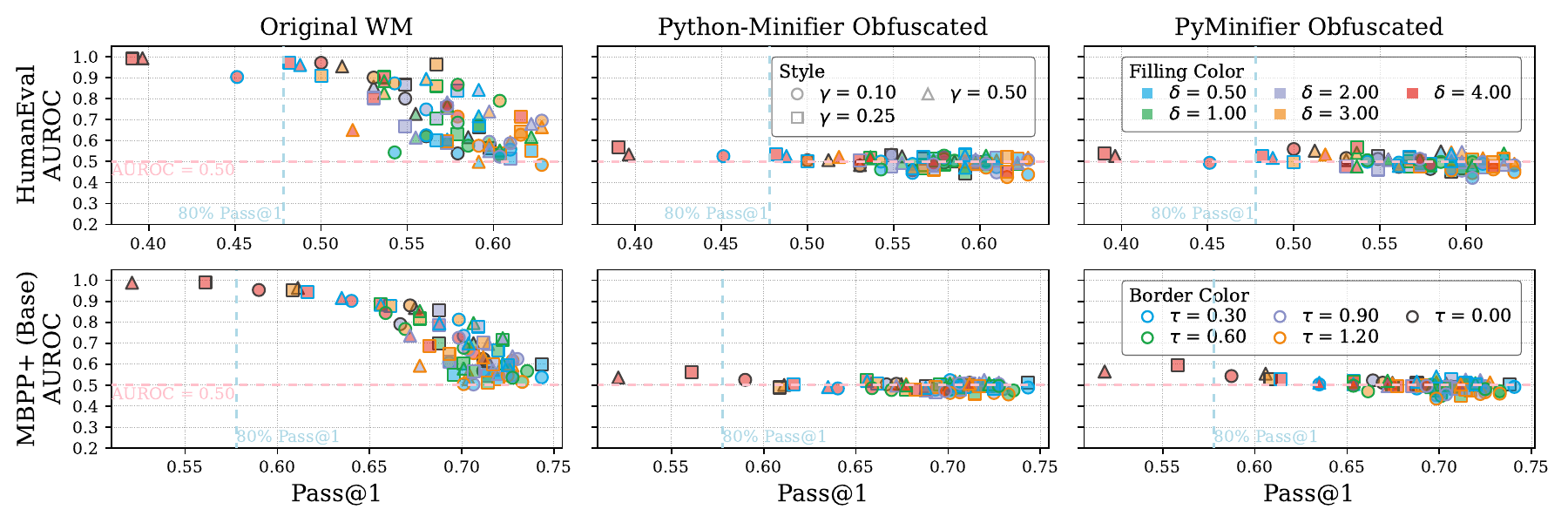}\\
    \vspace{0.0em}
    \includegraphics[width=\textwidth, trim={0 2.7mm 0 0}, clip]{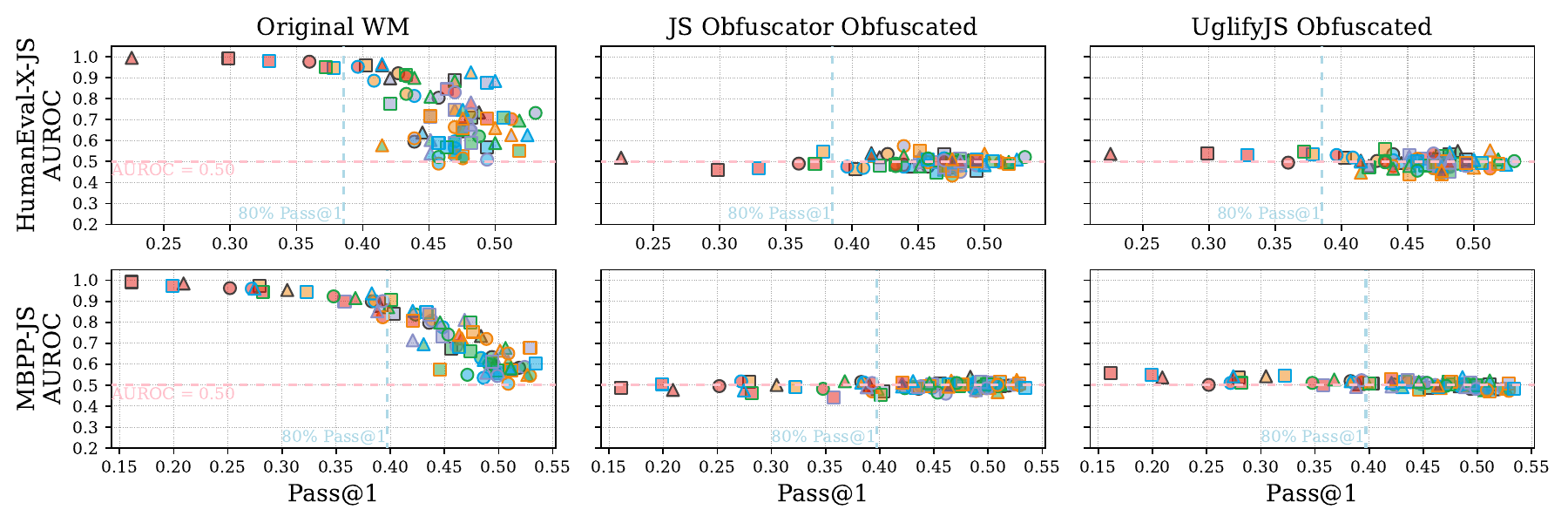}
    \caption{WLLM and SWEET \cite{kirchenbauer2023watermark, lee2023wrote} on LLaMA-3.1-8B-Instruct \cite{grattafiori2024llama}, before and after attack. The upper panel shows the effect of our attack on Python, and the lower panel shows the effect of our attack on JavaScript. The y-axis is AUROC and the x-axis is Pass@1; colors and markers denote hyperparameters. For DeepSeek results, see \cref{fig:green_red_app}.}
    \label{fig:green_red_main}
\end{figure*}
\begin{figure}[t]
    \centering
    \includegraphics[width=1\linewidth, trim={0 0 0 0}, clip]{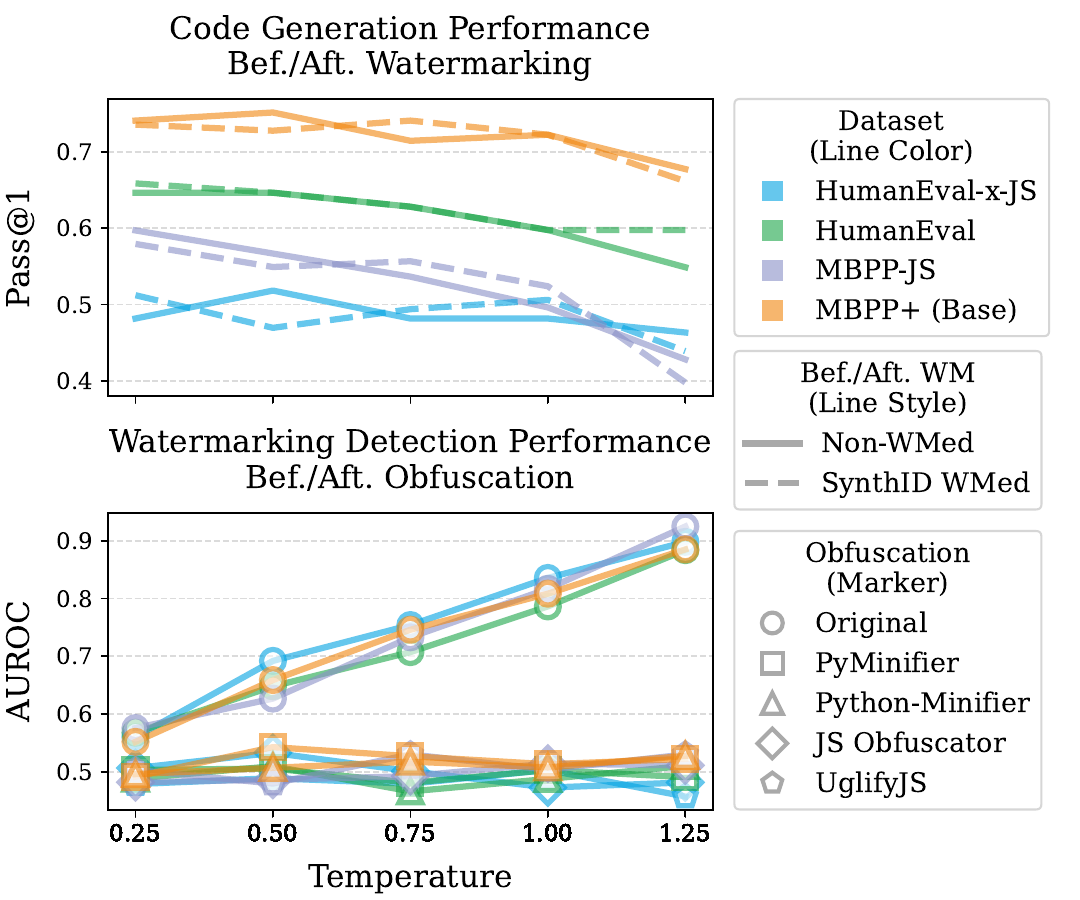}
    \caption{
    LLaMA 3.1 Pass@1 versus temperature with and without SynthID \cite{dathathri2024scalable}, and detection AUROC versus temperature before and after obfuscation. See \cref{fig:synthid_deepseek} for DeepSeek results.
    }\label{fig:synthid_llama}
\end{figure}

{
\begin{figure}[!b]
    \centering
    \includegraphics[width=\columnwidth, trim={0 3.3mm 0 0}, clip]{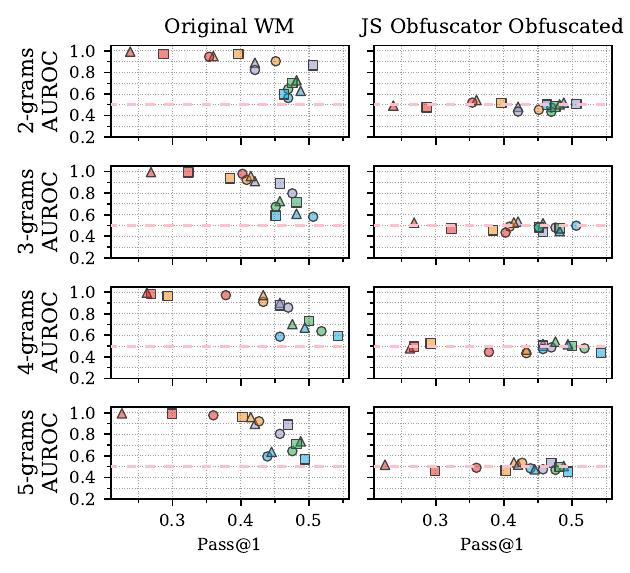}

    \caption{
    WLLM \cite{kirchenbauer2023watermark} under 2- to 5-gram settings on LLaMA 3.1 \cite{grattafiori2024llama}, before and after attack, on HumanEval-X-JS \cite{zheng2023codegeex}. Left columns show original AUROC-Pass@1 trade-offs; right columns show attacked results.}
    \label{fig:ngram_len}
\end{figure}
}

In this section, to support our theory, we implement our attack scheme against three state-of-the-art watermarking schemes.
We use off-the-shelf code obfuscators to conduct a low-cost, practical attack.
Code obfuscators generally integrate various equivalent transformation rules to perturb input code.
Compared with the randomized setting in our theoretical analysis, some realistic obfuscators apply fixed transformation rules or produce a fixed output for each input.
Without losing independence from N-gram features, such obfuscators can be viewed as implementations of our attack algorithm with a fixed random seed.
\cref{sec:rw_obf} additionally evaluates a randomized obfuscator that directly implements the reversible random-walk construction; it drives all effective settings to the same chance-level detection.

\subsection{Experiment Setting}
\label{sec:experiment_setting}

\textbf{Models and watermarking schemes.}
We use two LLMs with strong code-generation capabilities: LLaMA-3.1-8B-Instruct~\cite{grattafiori2024llama, meta2024llama} and DeepSeek-Coder-33B-Base~\cite{guo2024deepseek}.
We evaluate three state-of-the-art watermarking schemes: SWEET~\cite{lee2023wrote}, WLLM~\cite{kirchenbauer2023watermark}, and SynthID~\cite{dathathri2024scalable}.
SWEET is code-specific, whereas WLLM and SynthID are representative distortionary and non-distortionary N-gram watermarking schemes.

For WLLM and SWEET, we sweep green-token rates $\gamma \in \{0.1,0.25,0.5\}$ and logit biases $\delta \in \{0.5,1.0,2.0,3.0,4.0\}$.
SWEET additionally uses entropy thresholds $\tau \in \{0.3,0.6,0.9,1.2\}$.
For SynthID, we use $m=30$ tournament rounds with two competitors, and vary temperature over $\{0.25,0.5,0.75,1.0,1.25\}$.
Following the original papers, these values cover typical hyperparameter settings for each scheme.
Unless otherwise stated, all schemes use 5-grams, and all other LLM generation parameters use their default values.

\textbf{Benchmarks and metrics.}
For Python, we evaluate code generation on HumanEval~\cite{chen2021evaluating} and MBPP+ under the \textit{base} setting~\cite{liu2023is}.
For JavaScript, we use HumanEval-X-JS~\cite{zheng2023codegeex} and MBPP-JS from MultiPL-E~\cite{cassano2022multipl}.
Code quality is measured by Pass@1~\cite{chen2021evaluating}; watermark detectability is measured by AUROC between the $z$-score distributions of the generated samples and the standard normal distribution.
Uncompilable samples are excluded from AUROC because obfuscation is only meaningful for compilable code.
The overall compilation success rate of watermarked samples is $97.0\%$, indicating minimal impact on the statistical validity of our results.

\textbf{Obfuscators.}
For Python, we use Python-Minifier~\cite{python_minifier_repo} and PyMinifier~\cite{pyminifier_repo}; for JavaScript, we use JS Obfuscator~\cite{javascript_obfuscator_repo} and UglifyJS~\cite{uglifyjs_repo}.
These obfuscators are designed to preserve semantic equivalence, aiming to keep the code's functionality and correctness intact after obfuscation.
In our experiments, Python-Minifier, JS Obfuscator, and UglifyJS preserve test passability, while PyMinifier causes a small $0.21\%$ failure rate in preserving passability.
Upon manual inspection of several failed cases, all issues stemmed from PyMinifier wrongly transforming single-element tuples into standalone elements, thereby altering the original semantics.

\subsection{SOTA Watermarking Schemes Under Attack}
\label{sec:main_exp_result}
\label{sec:detection_perform}

\cref{fig:green_red_main} and \cref{fig:green_red_app} show the WLLM/SWEET results on LLaMA 3.1 and DeepSeek-Coder.
\cref{fig:synthid_llama} and \cref{fig:synthid_deepseek} show the results for SynthID.
Before the attack, the tested watermarking schemes preserve generation quality while achieving high detectability.
On LLaMA 3.1, all three schemes have hyperparameter settings with AUROC above $0.9$ while retaining more than $80\%$ of the corresponding non-watermarked Pass@1.
On DeepSeek-Coder, SynthID reaches AUROC $0.9$ at temperature $1.25$ with less than a $2\%$ Pass@1 drop.
See Appendix \ref{src:original_performance} for detailed analysis.

After obfuscating the watermarked code, detection collapses.
Under the distribution-consistency assumption in \cref{asm:main}, obfuscated watermarked code should have N-gram statistics close to ordinary code, so the detector should approach random guessing, i.e., AUROC $0.5$.
This is exactly what we observe in the results (Figures \ref{fig:green_red_main}, \ref{fig:synthid_llama}, \ref{fig:green_red_app}, and \ref{fig:synthid_deepseek}).
Across models, languages, benchmarks, and watermarking schemes, post-attack AUROC values concentrate tightly around $0.5$.
For each language, one obfuscator makes \emph{all} post-attack AUROC scores for \emph{all} three watermarking schemes fall within $(0.4,0.6)$ (Python-Minifier and JS Obfuscator).
Even weaker or more naive obfuscators largely nullify detection: $99.7\%$ of UglifyJS results fall in this range, as well as $97.5\%$ of results under PyMinifier, which only minifies code without further perturbations.
Overall, only $9/1280=0.7\%$ of obfuscated data points lie outside $(0.4,0.6)$, and none of these preserve more than $80\%$ of the default non-watermarked Pass@1.
Thus, \emph{every} configuration that preserves $\geq 80\%$ of the default non-watermarked Pass@1 becomes effectively undetectable after obfuscation.
The collapse persists on repository-level tasks from SWE-bench Verified (\cref{sec:swebench}), and when AUROC is computed against four concrete negative classes instead of the standard normal reference (\cref{sec:neg_class}).

\subsection{Impact of N-gram Length}
\label{sec:ngram_len}

As shown in \cref{fig:ngram_len}, we further test whether the attack depends on the chosen N-gram length.  Using WLLM, JS Obfuscator, and LLaMA 3.1, we vary the N-gram length from 2 to 5. 
Before attack, WLLM is relatively insensitive to this choice: the average Pass@1 values are $(0.419,0.426,0.429,0.420)$, and the average AUROC values are $(0.812,0.818,0.824,0.819)$ for 2- through 5-grams.  
After the attack, our conclusion remains unchanged. All post-obfuscation AUROC scores cluster around $0.5$, and no data point exceeds $0.6$.
This suggests that changing the N-gram order does not restore robustness.

\subsection{Distribution Consistency Test}
\label{sec:dist_test_ext}

Finally, we directly test the core assumption behind our theory: distribution consistency (\cref{asm:main}).
On all three watermarking schemes from our experiments, we use a theory-grounded random-walk obfuscator (\cref{sec:rw_obf}) to sample within equivalent spaces.
As seed programs, we take all highly watermarked programs (z-score $> 4$) from our experiments that pass all HumanEval or MBPP test cases.
This yields $1{,}620$ Python programs with an average z-score of $5.9$.
For each seed program, we run $30$ independent $100$-step random walks; since the walk operates within the seed's rule-based equivalent space, this constitutes random sampling within that space.
We then apply the Anderson--Darling goodness-of-fit test \cite{stephens1974edf} to each equivalent space, with the null hypothesis $H_0$ that the $z$-scores within the space follow $\normal(0,1)$, the standard normal distribution of the broader code space---that is, that distribution consistency holds for the space.

\begin{table}[H]
    \centering
    \small
    \begin{tabular}{l r r r}
        \specialrule{1.2pt}{0pt}{0pt}
        Scheme & Spaces & $H_0$ not rejected & Rate \\
        \specialrule{0.9pt}{0pt}{0pt}
        SWEET & 867 & 827 & 95.39\% \\
        WLLM & 700 & 678 & 96.86\% \\
        SynthID & 53 & 52 & 98.11\% \\
        \hline
        Overall & 1,620 & 1,557 & 96.11\% \\
        \specialrule{1.2pt}{0pt}{0pt}
    \end{tabular}
    \caption{
    Anderson--Darling consistency test over $1{,}620$ equivalent spaces sampled with the random-walk obfuscator. For each scheme, we report the number of equivalent spaces whose internal z-score distribution is not rejected as standard normal at the $5\%$ significance level.
    }
    \label{tab:dist_test_ext}
\end{table}

\cref{tab:dist_test_ext} summarizes the outcome.
At the $5\%$ significance level, the test does not reject $H_0$ for $1{,}557$ of the $1{,}620$ equivalent spaces ($96.11\%$), with per-scheme rates between $95.39\%$ and $98.11\%$.
Across all three watermarking schemes, a practical rule set can satisfy distribution consistency to a high degree, leaving little room for the watermark signal to survive within the equivalent space.






\section{Conclusion}

We show that N-gram-based code watermarking is fundamentally fragile under semantics-preserving obfuscation. Our theory models obfuscation as a Markov random walk and proves that, under the experimentally supported assumption of distribution consistency, robustness collapses: after enough obfuscation, detection fails at rate $1-\fpr$.

Experiments across schemes, models, languages, benchmarks, and obfuscators confirm this prediction: detection degrades to near-random performance. This suggests that surface-form N-gram signals are insufficient for robust code watermarking, motivating semantics-aware designs.
\section*{Limitations}




\paragraph{Attack on Other Watermarking Schemes}
Recently, several watermarking schemes have been proposed that avoid the N-gram-based routine to improve robustness \cite{zhao2023provable, liu2024adaptive}, while this work is limited to N-gram-based schemes. 

However, given that N-gram-based watermarking is both widely adopted \cite{dathathri2024scalable, kirchenbauer2023watermark, lee2023wrote, kirchenbauer2023reliability, liu2023unforgeable, wu2023resilient, zhao2024permute, fu2024gumbelsoft} and industrially deployed \cite{dathathri2024scalable}, we believe that the robustness analysis of N-gram-based watermarking remains timely and valuable.

In addition, some studies inject watermarks using code transformation rules during post-processing or code LLM training dataset preparation \cite{yang2024srcmarker, sun2023codemark, li2024resilient}. 
These approaches, as well as other software watermarking schemes \cite{dey2019software}, often rely on language-specific transformation rules.
Although we focus on N-gram-based watermarking schemes, our attack would also disrupt such watermarks, since the transformation rules can be reversed directly by obfuscators, depending solely on whether the obfuscator implements them.

\paragraph{Different Code Quality Aspects}
This work evaluates code quality primarily through execution behavior, measured by whether generated programs pass test cases.
Specifically, we use code benchmarks to compute the Pass@1 metric, following a widely adopted evaluation practice \cite{chen2021evaluating, liu2023is, zheng2023codegeex, cassano2022multipl}.
We acknowledge that this does not cover all dimensions of code quality, such as readability, maintainability, or efficiency.
Nevertheless, execution behavior is arguably the most fundamental dimension: its prominence in widely used code benchmarks reflects the fact that functional correctness is often the primary criterion for assessing generated code. In some settings, such as programming competitions, obfuscated binary deployment, and black-box API consumption, it may even be the main or only aspect of concern.
In the future, developing obfuscation techniques that retain multiple dimensions of code quality could be a promising direction.

\section*{Ethical Considerations}

\paragraph{Data sources and licensing.}
Our experiments use public code-generation benchmarks and open-source obfuscation tools, including HumanEval, MBPP+, HumanEval-X-JS, MBPP-JS from MultiPL-E, and Python and JavaScript obfuscators. We use these resources only for research and evaluation, and we follow their original licenses and terms of use. Our released repository is intended to contain only our own code, evaluation scripts, and links or instructions for obtaining third-party resources from their original sources; we do not claim ownership over benchmark data, model outputs, or external software distributed by others.

\paragraph{Privacy and sensitive information.}
The datasets used in this work are benchmark programming tasks and tests rather than human-subject or user-interaction data, so we do not intentionally collect personal data or sensitive user information. We also do not deploy the studied watermarking or attack methods on real users, production systems, or private codebases.

\section*{Acknowledgments}
This material is based upon work supported by the National Science Foundation under Grant Nos.~2342250, 2319944, and 2542053.

Research reported in this publication was supported by an Amazon Research Award, 2024, and an NVIDIA Academic Grant Program award, 2025.

We use AI writing assistance solely to improve the language of the manuscript, including paraphrasing or polishing the authors' original text, rather than suggesting new content.


\bibliography{main}

\appendix

\section{N-gram-based Watermark}
\label{sec:ngram_detail}
In the discourse on tracing and regulating AI-generated content (AIGC), watermarking offers a promising solution. Among various approaches, those that modify the output probability distribution of a language model during generation are of particular interest due to their model-agnostic nature. Following the foundational framework proposed by \cite{kirchenbauer2023watermark}, which embeds a signal by modifying vocabulary logits, we provide a detailed definition and workflow of the N-gram-based watermarking technique. This process is divided into two core stages: \textbf{Watermark Embedding} and \textbf{Watermark Detection}.

\subsection{Watermark Embedding}
Watermark embedding occurs at every token generation step of the Large Language Model (LLM). The objective is to slightly alter the sampling probability distribution of the next token to favor a verifiable pattern, without significantly degrading the quality of the text.

\noindent\textbf{Context-based Pseudorandom Seed Generation.}
During the generation of the $i$-th token, $t_i$, the watermarking algorithm uses the preceding $N-1$ tokens $(t_{i-N+1}, \dots, t_{i-1})$ as context, so that the context together with the current token spans an N-gram. This context is used to generate a deterministic, pseudorandom seed.

A seed $s_i$ is generated using a public hash function $H$ (e.g., SHA-256):
$$s_i = H(t_{i-N+1}, \dots, t_{i-1}).$$
To enhance security, a secret key $k$ known only to the owner can be incorporated: $s_i = H(t_{i-N+1}, \dots, t_{i-1}, k)$. In the simplest case, $N=2$, the context is the single preceding token $t_{i-1}$.
The first $N-1$ tokens of a generation are sampled from the unmodified distribution, since no full $(N-1)$-token context is available for them.

\noindent\textbf{Vocabulary Partitioning.}
After obtaining the seed $s_i$, it is used to initialize a Pseudorandom Number Generator (PRNG), which partitions the entire vocabulary $V$ into two disjoint sets in a reproducible manner: \textbf{Green List} ($G_i$) and \textbf{Red List} ($R_i$).

The partition is governed by a hyperparameter $\gamma \in (0, 1)$, which represents the fraction of the vocabulary assigned to the green list.
$$V = G_i \cup R_i \quad \text{and} \quad G_i \cap R_i = \emptyset,$$
$$|G_i| \approx \gamma |V|.$$
Since $s_i$ is determined by the context $(t_{i-N+1}, \dots, t_{i-1})$, the contents of the green list $G_i$ and red list $R_i$ are identical for the same context.

\noindent\textbf{Logit Modification.}
After the LLM generates the original logits vector $l_i = (l_{i,1}, l_{i,2}, ..., l_{i,|V|})$ for position $i$, the watermarking algorithm modifies it. Specifically, it adds a fixed bias $\delta > 0$ to the logits of all tokens in the green list $G_i$.

The modified logits vector $l'_i$ is computed as follows:
$$l'_{i,v} = \begin{cases} l_{i,v} + \delta & \text{if token } v \in G_i, \\ l_{i,v} & \text{if token } v \in R_i. \end{cases}$$
This operation significantly increases the probability of sampling tokens from the green list. Finally, the model samples the $i$-th token $t_i$ from this modified probability distribution:
$$t_i \sim \text{Softmax}(l'_i).$$

\subsection{Watermark Detection}

The detection process aims to determine if a given text $T = (t_1, t_2, ..., t_L)$ contains a watermark.

\noindent\textbf{Reconstructing the Green Lists.}
The detector possesses the same hash function $H$ (and secret key $k$) used during embedding. For each token $t_i$ in the text (from $i=N$ to $L$), the detector uses the preceding $N-1$ tokens $(t_{i-N+1}, \dots, t_{i-1})$ to regenerate the exact green list $G_i$ that was used at the time of generation. The first $N-1$ tokens, which lack a full context, are not scored.

\noindent\textbf{Hypothesis Testing.}
The detection process is a standard statistical hypothesis test:
\begin{itemize}
    \item \textbf{Null Hypothesis ($H_0$)}: The text was not generated by a watermarked LLM. Therefore, the probability of any token $t_i$ appearing in its corresponding green list $G_i$ is approximately equal to the green list proportion, $\gamma$.
    \item \textbf{Alternative Hypothesis ($H_A$)}: The text was generated by a watermarked LLM. Due to the logit modification, the probability of a token $t_i$ appearing in its green list $G_i$ is significantly greater than $\gamma$.
\end{itemize}

\noindent\textbf{Calculating the Test Statistic.}
First, among the $m = L-N+1$ scored positions, we count the total number of tokens $K$ that fall into their corresponding green lists:
$$K = \sum_{i=N}^{L} \mathbb{I}(t_i \in G_i).$$
where $\mathbb{I}(\cdot)$ is the indicator function (1 if the condition is true, 0 otherwise).

Under the null hypothesis $H_0$, the count $K$ should follow a binomial distribution $B(m, \gamma)$. For a sufficiently long text (large $m$), this binomial distribution can be approximated by a normal distribution.

We can then compute a \textbf{z-score} to measure how much the observed count $K$ deviates from the expected count:
$$z = \frac{K - \mu}{\sigma} = \frac{K - m\gamma}{\sqrt{m\gamma(1-\gamma)}}.$$
where:
\begin{itemize}
    \item $K$ is the observed count of green-list tokens.
    \item $\mu = m\gamma$ is the expected value of $K$ under $H_0$.
    \item $\sigma = \sqrt{m\gamma(1-\gamma)}$ is the standard deviation of $K$ under $H_0$.
\end{itemize}
A sufficiently large positive z-score (e.g., $z > 4$ or $5$) provides strong statistical evidence to reject the null hypothesis $H_0$, leading to the conclusion that the text is watermarked.

\section{Notion Formalizations}
\label{apx:sec:theory_details}

In this section, we formalize the main concepts used in our theory, partly following prior definitions in \cite{zhang2023watermarks, zhao2024sok}.

\subsection{Threat Model}
\label{apx:sec:threat_model}
Let $\promptSpace$ be a prompt space, $\codeSpace$ a code space, $\model:\promptSpace\to\codeSpace$ a randomized code generation model, and $\testSuite:\promptSpace\times\codeSpace\to \curly{0,1}$ a code test suite that checks whether a candidate program satisfies a prompt. A secret-key watermarking scheme $\wmScheme=(\watermark, \detect)$ consists of two algorithms:
\begin{itemize}
    \subdef{$\watermark(\model)$}
        Given $\model \in \modelSpace$, this randomized algorithm outputs a secret key $\secretKey \in \keySpace$ and a watermarked model $\model_{\secretKey}: \promptSpace \to \codeSpace$ parameterized by $\secretKey$.
    \subdef{$\detect_{\secretKey}(\prompt,\code)$}
        Given a secret key $\secretKey \in \keySpace$, a prompt $\prompt \in \promptSpace$, and a code sample $\code \in \codeSpace$, this deterministic detector outputs $\detect_{\secretKey}(\prompt, \code)\in\curly{0,1}$, where $1$ indicates that the watermark is present and $0$ indicates that it is absent.
\end{itemize}

The goal is that $\model_{\secretKey}$ generates watermarked code that can later be recognized by $\detect_{\secretKey}$. The secret key $\secretKey$ may be chosen and maintained by the model provider or by another designated party.

We evaluate the detector using the false negative rate (FNR)
\[
    \fnr \gets \expect(\indicator[\detect_{\secretKey}(\prompt,\code)=0] \mid \code \in \codeSpace_{\secretKey})
\]
and the false positive rate (FPR)
\[
    \fpr \gets \expect(\indicator[\detect_{\secretKey}(\prompt,\code)=1] \mid \code \in \codeSpace),
\]
where $\codeSpace_{\secretKey} \subseteq \codeSpace$ denotes the set of watermarked responses produced by $\model_{\secretKey}$. For a usable watermarking scheme, both $\fpr$ and $\fnr$ should be sufficiently small.

For a prompt $\prompt$, let $\func$ denote the behavior requested by the prompt, and let $\hqSpace \subseteq \codeSpace$ denote the \textit{high-quality space}, namely the set of programs $\code$ such that $\testSuite(\prompt, \code)=1$. Code quality can be measured along many dimensions, but in this work both $\func$ and $\testSuite$ refer only to runtime behavior, such as correctness and functional compliance. This focus follows standard practice in code-generation benchmarks, where execution behavior is the primary quality criterion \cite{chen2021evaluating, liu2023is, zheng2023codegeex, cassano2022multipl}.

\textbf{Adversary Attacker:} An adversary $\adversary: \promptSpace \times \codeSpace \to \codeSpace$ acts as a code modifier or obfuscator and satisfies
\[
    \forall \code \in \hqSpace, \adversary(\prompt, \code) \in \hqSpace.
\]
Equivalently, $\testSuite(\prompt, \adversary(\prompt, \code))=1$ whenever $\testSuite(\prompt, \code)=1$. Thus, given a prompt $\prompt$ and a high-quality program $\code$, the attacker outputs a transformed program $\code' \gets \adversary(\prompt, \code)$ without degrading execution quality.

We use $\aRate$-breaking to formalize the attack goal. An attacker $\aRate$-breaks the watermarking scheme if, for every $\model \in \modelSpace$ and every prompt $\prompt \in \promptSpace$, we have
\[
    \expect(\indicator[\detect_{\secretKey}(\prompt, \adversary(\prompt, \code))=0] \mid \code \in \codeSpace_{\secretKey}) \geq \aRate,
\]
that is, after the quality-preserving transformation by $\adversary$, the resulting false negative rate is at least $\aRate$.

From the defender's perspective, a watermarking scheme is robust if every feasible attacker can only force $\aRate$ to remain close to the baseline $\fnr$. In that case, the detector continues to work even after adversarial code modifications.

\section{Equivalent Transformations}
\label{apx:sec:eq_trans}

In this section, we formalize equivalent transformation rules and the rule-based equivalent spaces they induce. The general idea of changing program syntax while preserving behavior is standard in compiler optimization, refactoring, and code obfuscation \cite{behera2015different, banescu2018tutorial, peng2019adaptive, shah2018code}. Our attack model builds on this idea: it treats semantics-preserving transformations as the moves of a random walk over high-quality programs.

\begin{definition}[Equivalent Transformation Rule and Executor]
\label{apx:def:transformation_rule}
    Given a program $\code \in \codeSpace$, we use two algorithms to model equivalent transformations:
    \begin{itemize}
        \subdef{$\eqRule(\code)$}
            Each rule $\eqRule$ is an action-derivation algorithm. Given $\code$, it returns a set of executable actions,
            \[
                \eqRule: \codeSpace \to \bigcup_{i=1}^\infty \actionSpace^i,
            \]
            that can be applied to $\code$ by $\trans(,)$ below. Different rules in the rule space $\ruleSpace$ correspond to different transformation types. Here $\action \in \actionSpace$ denotes one concrete transformation action.
        \subdef{$\trans(\code, \action)$}
            This executor is shared across rules. Given a program and an action $\action \in \actionSpace$ compatible with the current rule, $\trans: \codeSpace \times \actionSpace \to \codeSpace$ performs a quality-preserving transformation satisfying
            \[
                \forall \func, \forall \code \in \hqSpace, \trans(\code, \action) \in \hqSpace,
            \]
            where $\action$ must be one of the actions allowed by the rule currently under consideration.
    \end{itemize}
\end{definition}

The point of this abstraction is to separate \emph{what kind of transformation} is allowed from \emph{which concrete modification} is executed. A rule describes a transformation family, whereas $\trans$ applies one specific action from that family.

For example, one rule may correspond to ``delete a comment,'' in which each returned action deletes one specific comment at one specific location in $\code$. Another rule may correspond to ``rewrite a \texttt{for}-loop as a \texttt{while}-loop,'' in which each action targets one concrete loop instance. Many such quality-preserving rules are readily implementable in practice, including comment removal, variable renaming, and dead-code insertion \cite{python_minifier_repo,pyminifier_repo,javascript_obfuscator_repo,uglifyjs_repo}.

We now define the randomized transformation procedure used in our attack.

{
\newcommand{\oriCode}{\code_{\mathsf{ori}}}
\newcommand{\transCode}{\code_{\mathsf{trans}}}
\newcommand{\randomChoose}{\mathsf{RandomChoose}}

\begin{definition}[Random Transformation]
\label{apx:def:random_transformation}
    Let $\randomChoose: \bigcup_{i=1}^\infty \nature^i \to \nature$ be an algorithm that samples one integer uniformly from a finite integer set. Using $\randomChoose$, we define two randomized algorithms, $\step(\code, \ruleSet)$ and $\walk(\code, \ruleSet, \stepNum)$, in Algorithms \ref{apx:alg:random_step} and \ref{apx:alg:random_walk}.
\end{definition}

\begin{algorithm}[tbp]
\caption{$\step$}
\label{apx:alg:random_step}
    \KwIn{
        Original Code: $\oriCode \in \codeSpace$,  \newline
        Rule Set: $\ruleSet=\curly{\eqRule_1, \eqRule_2, \dots, \eqRule_n}$
    } \KwOut {
        Transformed Code: $\transCode \in \codeSpace$
    }
    Init Action Set $\actionSet \gets \varnothing$ \\
    \For{each $i$  in $1,2, \dots, n$} {
        Set $\actionSet \gets \actionSet \cup \eqRule_i(\oriCode)$
    }
    Init $j \gets \randomChoose(\curly{1,2,\dots, \sizeof{\actionSet}})$ {\small \tcp{$\actionSet$  is $\curly{ \action_1, \action_2, \dots, \action_{\sizeof{\actionSet}}}$, and $\actionSet \subseteq \actionSpace$}}
    \label{apx:line:choose}
    Init $\transCode \gets \trans(\oriCode, \action_j)$ \\
    Return $\transCode$
\end{algorithm}

The algorithm $\step$ (\cref{apx:alg:random_step}) performs one quality-preserving transformation on the current program. It first aggregates all actions enabled by the rule set $\ruleSet$ and then selects one action uniformly at random from the resulting set $\actionSet$ (\cref{apx:line:choose}). The algorithm $\walk$ applies $\step(\code, \ruleSet)$ repeatedly for a finite number of steps $\stepNum$.

Importantly, in our model, neither $\step$ nor $\walk$ depends on the prompt or on prompt-specific semantic annotations. In software engineering practice, many semantics-preserving transformations can be carried out directly from the program text once the language grammar is known.

We next define the two central notions used in the analysis: ergodicity rule sets and rule-based equivalent spaces.

\begin{definition}[Ergodicity Rule Set]
\label{apx:def:erg_rule_set}
    An ergodicity rule set $\ergRuleSet = \curly{\eqRule_1, \eqRule_2, \dots, \eqRule_n}$ is a set of rules satisfying the following properties:
    \begin{itemize}
        \subdef{Empty Rule}
            The set contains a rule that performs empty transformations. That is, for all $\prompt \in \promptSpace$, $\code \in \codeSpace$, and $\action \in \eqRule(\code)$, the corresponding executor satisfies
            \[
                \code = \trans(\code, \action).
            \]
        \subdef{Inverse Rule}
            Every rule $\eqRule \in \ergRuleSet$ has an inverse rule $\eqRule^{-1}$ such that, if $\actionSet = \eqRule(\code)$ and $\code' = \trans(\code, \action)$, then for each action $\action \in \actionSet$ there exists at least one inverse action $\action^{-1} \in \eqRule^{-1}(\code')$ satisfying
            \[
                \code = \trans(\code', \action^{-1}).
            \]
            Consequently, every multi-step transformation is reversible in the reachability sense. For brevity, write
            \begin{gather}
                p_{\stepNum}(\code_i, \code_j) \\
                := \\
                \Pr[\code_j = \walk(\code_i, \ergRuleSet, \stepNum)].
            \end{gather}
            Then, for all $\code_{+}, \code_{-} \in \codeSpace$ and all $\stepNum \in \nature$,
            \[
                p_{\stepNum}(\code_{+}, \code_{-}) > 0
                \iff
                p_{\stepNum}(\code_{-}, \code_{+}) > 0.
            \]
    \end{itemize}
\end{definition}

{
\newcommand{\rom}[1]{\textsc{\romannumeral #1}}
\begin{table}[b]
    \centering
    \small
    \begin{tabularx}{\columnwidth}{@{}c c >{\raggedright\arraybackslash}X@{}}
        \specialrule{1.2pt}{0pt}{0pt}
        $\eqRule$ & $\eqRule^{-1}$ & \textsc{Description} \\
        \specialrule{0.9pt}{0pt}{0pt}
        \rom{1} & \rom{1} & Do nothing. \\ \hline
        \rom{2} & \rom{2} & Randomly rename variables. \\ \hline
        \rom{3} & \rom{4} & Add a random comment. \\ \hline
        \rom{4} & \rom{3} & Delete a comment. \\ \hline
        \rom{5} & \rom{6} & Add a random dead code snippet. \\ \hline
        \rom{6} & \rom{5} & Delete a dead code snippet.  \\
        \specialrule{1.2pt}{0pt}{0pt}
    \end{tabularx}
    \caption{Example of an ergodicity rule set. The left column lists $\eqRule$ and the middle column lists $\eqRule^{-1}$.}
    \label{apx:tab:erg_set_example}
\end{table}

An ergodicity rule set is not merely a theoretical construct; it can be implemented using standard code transformations. \Cref{apx:tab:erg_set_example} gives one example in which each rule has a corresponding inverse rule. Some rules can even be self-inverse. For instance, although reversing comment deletion by randomly adding a comment may have low probability, that probability is still nonzero, which is sufficient for \cref{apx:def:erg_rule_set}.
}

\begin{definition}[Rule-based Equivalent Space]
\label{apx:def:rule_based_eq_space}
    Given a prompt $\prompt$, an initial code sample $\code \in \hqSpace$, a rule set $\ruleSet = \curly{\eqRule_1, \dots, \eqRule_n}$, and an arbitrary finite step count $\stepNum \in \nature$, the rule-based equivalent space $\eqSpace^{\ruleSet}_{\code}$ consists of the initial code $\code$ and all programs reachable as $\walk(\code, \ruleSet, \stepNum)$. In particular, $\eqSpace^{\ruleSet}_{\code} \subseteq \hqSpace \subseteq \codeSpace$.
\end{definition}

Equivalently, $\eqSpace^{\ruleSet}_{\code}$ collects every high-quality program that can be reached from $\code$ by applying the transformations in $\ruleSet$ for any finite number of steps.

\subsection{Chain Modeling}
\label{apx:sec:chain_modeling}

We now model rule-based code transformation as a graph, because our attacker operates as the random walk defined in \cref{apx:def:random_transformation}.

\begin{algorithm}[tbp]
\caption{$\walk$}
\label{apx:alg:random_walk}
    \KwIn{
        Original Code: $\oriCode \in \codeSpace$, \newline
        Rule Set: $\ruleSet=\curly{\eqRule_1, \eqRule_2, \dots, \eqRule_n}$,  \newline
        Number of Steps: $\stepNum \in \nature$
    } \KwOut{
        Transformed Code: $\transCode \in \codeSpace$
    }
    Init $\code \gets \oriCode$ \\
    \For{each $i$ in $1,2, \dots, \stepNum$}{
        Set $\code \gets \step(\code, \ruleSet)$
        \label{apx:line:call_step}\\
    }
    Init $\transCode \gets \code$ \\
    Return $\transCode$
\end{algorithm}
}

Given an initial program $\code \in \codeSpace$ and a rule set $\ruleSet$, the induced rule-based equivalent space $\eqSpace^{\ruleSet}_{\code}$ gives rise to a weighted directed graph $\graph^{\ruleSet}_{\code}=(\vertexSet,\edgeSet,\weightSet)$ describing the possible transformations inside that space. If $\eqSpace^{\ruleSet}_{\code} = \curly{\code_1, \code_2, \dots, \code_n}$, then each vertex $\vertex_i \in \vertexSet$ corresponds to program $\code_i$. An edge $\edge_{ij} \in \edgeSet$ exists whenever
\[
    \Pr[\code_j = \step(\code_i, \ruleSet)] > 0,
\]
meaning that $\code_i$ can be transformed into $\code_j$ in one step. Each such edge carries weight $\weight_{ij} = \Pr[\code_j = \step(\code_i, \ruleSet)] \in (0,1]$.

Because \cref{apx:alg:random_step} chooses uniformly from the available action set, all outgoing edges from the same vertex have equal weight. Formally,
\[
    \forall \weight_{i+}, \weight_{i-} \in \curly{\weight_{ij} \mid \edge_{ij} \in \edgeSet, \forall j},
    \quad
    \weight_{i+} = \weight_{i-}.
\]

Based on $\graph^{\ruleSet}_{\code}$, we define the transition matrix $\transMat \in \real^{\sizeof{\vertexSet} \times \sizeof{\vertexSet}}$ by
\[
    \transMat_{ij} =
        \begin{cases}
        \weight_{ij}, & \edge_{ij} \in \edgeSet, \\
        0, & \text{otherwise}.
        \end{cases}
\]

For each starting program $\code_i$, define the probability distribution vector $\distVec(\code_i, \stepNum) \in \real^{\sizeof{\vertexSet}}$ by
\[
    \distVec(\code_i, \stepNum)_j = \Pr[\code_j = \walk(\code_i, \ruleSet, \stepNum)].
\]
Thus, $\distVec(\code_i, \stepNum)_j \in [0,1]$ is the probability that the random walk, started from $\code_i$, is at $\code_j$ after $\stepNum$ calls to $\step$ (\cref{apx:line:call_step}). For example, $\distVec(\code_1, 0) = [1,0,0,\dots,0]$ represents the initial state concentrated on $\code_1$. By construction,
\[
    \distVec(\code, \stepNum+1) = \transMat^{\top} \cdot \distVec(\code, \stepNum).
\]

If $\transMat$ satisfies the standard conditions for the existence of a stationary distribution, namely irreducibility and aperiodicity of the corresponding Markov chain \cite{levin2017markov}, then there exists a stationary distribution $\statDist \in \real^{\sizeof{\vertexSet}}$ such that
\[
    \exists \statDist,
    \forall \code \in \eqSpace^{\ruleSet }_{\code},
    \lim_{\stepNum \to \infty}\distVec(\code, \stepNum)=\statDist.
\]

Following classical Markov-chain theory, we further define the mixing time
\[
    \stepNum_{i}(\aRate) = \min \curly{\stepNum \mid \tvDist{\distVec(\code_i, \stepNum)}{\statDist} \leq \aRate},
\]
which denotes the number of steps required for the walk started from $\code_i$ to come within total variation distance $\aRate$ of the stationary distribution $\statDist$.

\section{Model Analysis}

In this section, we analyze the structure induced by ergodicity rule sets. We first show in Appendix \ref{apx:sec:stationary_exist} that the corresponding code-transformation graph admits a stationary distribution.
Then, in Appendix \ref{apx:sec:rb_partition}, we prove that an ergodicity rule set induces a partition of the high-quality code space.

\subsection{The Stationary Distribution Exists}
\label{apx:sec:stationary_exist}

We now show that, inside the rule-based equivalent space induced by an ergodicity rule set, the random walk $\walk$ has a stationary distribution that does not depend on the starting program. By standard Markov-chain theory \cite{levin2017markov}, it suffices to prove irreducibility and aperiodicity.

\begin{theorem}[Inside-Space Independence]
\label{apx:tho:inside_ind}
    Let $\eqSpace^{\ergRuleSet }_{\code}$ be a rule-based equivalent space generated by an ergodicity rule set (\cref{apx:def:erg_rule_set}). Then
    \[
        \exists \statDist,
        \forall \code' \in \eqSpace^{\ergRuleSet }_{\code},
        \lim_{\stepNum \to \infty}\distVec(\code', \stepNum)=\statDist,
    \]
    where $\eqSpace^{\ergRuleSet }_{\code}$ is defined in Appendix \ref{apx:def:rule_based_eq_space} and $\distVec(,)$ is defined in Appendix \ref{apx:sec:chain_modeling}. Equivalently, the associated code-transformation graph $\graph^{\ergRuleSet}_{\code}$ is irreducible and aperiodic.
\end{theorem}

\begin{proof}[Proof of \cref{apx:tho:inside_ind}'s Irreducibility]
    By \cref{apx:def:rule_based_eq_space}, every program in $\eqSpace^{\ergRuleSet}_{\code}$ is reachable from the original program $\code$. Hence,
    \[
        \forall \code' \in \eqSpace^{\ergRuleSet}_{\code}, \exists \stepNum,
        \Pr[\code \to \code' \mid \stepNum] > 0,
    \]
    where $\Pr[\code \to \code' \mid \stepNum]$ abbreviates
    $\Pr[\code' = \walk(\code, \ergRuleSet, \stepNum)]$.

    Moreover, by the inverse-rule property in \cref{apx:def:erg_rule_set}, for all $\code_{+}, \code_{-} \in \codeSpace$,
    \[
    \begin{aligned}
        \Pr[\code_{+} \to \code_{-} \mid \stepNum] > 0
        \quad &\iff \quad \\
        \Pr[\code_{-} \to \code_{+} \mid \stepNum] > 0.
    \end{aligned}
    \]

    Now fix arbitrary $\code_{+}, \code_{-} \in \eqSpace^{\ergRuleSet}_{\code}$. Since both are reachable from $\code$, there exist $\stepNum_{+}, \stepNum_{-} \in \nature$ such that
    \[
        \Pr[\code \to \code_{+} \mid \stepNum_{+}] > 0,
        \qquad
        \Pr[\code \to \code_{-} \mid \stepNum_{-}] > 0.
    \]
    By reversibility of reachability, we also have
    \[
        \Pr[\code_{+} \to \code \mid \stepNum_{+}] > 0.
    \]
    Define
    \begin{gather*}
        p_{+} := \Pr[\code_{+} \to \code \mid \stepNum_{+}],
        \\
        p_{-} := \Pr[\code \to \code_{-} \mid \stepNum_{-}].
    \end{gather*}
    Then
    \[
        \Pr[\code_{+} \to \code_{-} \mid \stepNum_{+} + \stepNum_{-}]
        \geq p_{+} p_{-} > 0.
    \]
    Therefore, every pair of vertices in $\graph^{\ergRuleSet}_{\code}$ is mutually reachable through a finite path passing through $\code$. The graph is thus irreducible.
\end{proof}

\begin{proof}[Proof of \cref{apx:tho:inside_ind}'s Aperiodicity]
    By the empty-rule property in \cref{apx:def:erg_rule_set}, the graph contains a self-loop at every program with positive probability. In particular,
    \[
        \Pr[\code = \step(\code, \ergRuleSet)] > 0.
    \]
    Let
    \[
        p_{\mathrm{self}} := \Pr[\code = \step(\code, \ergRuleSet)].
    \]
    Hence, for every $\stepNum \in \nature$,
    \[
        \Pr[\code = \walk(\code, \ergRuleSet, \stepNum)]
        \geq p_{\mathrm{self}}^{\stepNum} > 0.
    \]
    Therefore, every positive integer belongs to the return-time set,
    \[
        \curly{\stepNum \mid \code = \walk(\code, \ruleSet, \stepNum)} = \nature,
    \]
    and thus its greatest common divisor is $1$. The graph is aperiodic.
\end{proof}

Note that $\eqSpace^{\ergRuleSet }_{\code}$ depends on both the rule set $\ergRuleSet$ and the initial program $\code$. Once these two objects are fixed, the corresponding space is uniquely determined.

Thus, \cref{apx:tho:inside_ind} states that within a fixed equivalent space, the long-run behavior of the walk is intrinsic to the space itself: regardless of the starting program, the distribution converges to the same stationary distribution $\statDist$.

\subsection{Rule-based Partition}
\label{apx:sec:rb_partition}

For the later impossibility result, we next show that an ergodicity rule set partitions the high-quality code space into disjoint equivalent spaces.

{
\newcommand{\setP}{\eqSpace^{\ergRuleSet }_{\code_{+}}}
\newcommand{\setN}{\eqSpace^{\ergRuleSet }_{\code_{-}}}
\newcommand{\setPSimp}{\eqSpace_{+}}
\newcommand{\setNSimp}{\eqSpace_{-}}
\begin{theorem}[Ergodicity Rule Set can Partition the High-Quality Code Space]
\label{apx:tho:partition}
    Given an ergodicity rule set $\ergRuleSet$, for all $\code_{+},\code_{-} \in \hqSpace$ we have
    \[
        \neg (\setP=\setN) \iff \setP \cap \setN = \varnothing
    \]
    where $\hqSpace \subseteq \codeSpace$. In other words, two distinct rule-based equivalent spaces do not overlap, so $\ergRuleSet$ induces a partition of the high-quality code space.
\end{theorem}

\begin{proof}[Proof of \cref{apx:tho:partition}]
    We abbreviate $\setP$ and $\setN$ as $\setPSimp$ and $\setNSimp$, and reuse the notation $\Pr[\to \mid]$ from the proof of \cref{apx:tho:inside_ind}. Assume
    \[
        \eqSpace := \setPSimp \cap \setNSimp \neq \varnothing.
    \]
    Let $\code \in \eqSpace$. Then for every $\code_{+} \in \setPSimp$, there exists $\stepNum_{+} \in \nature$ such that
    \[
        \Pr[\code_{+} \to \code \mid \stepNum_{+}] > 0.
    \]
    Likewise, for every $\code_{-} \in \setNSimp$, there exists $\stepNum_{-} \in \nature$ such that
    \[
        \Pr[\code \to \code_{-} \mid \stepNum_{-}] > 0.
    \]
    Define
    \begin{gather*}
        p_{+} := \Pr[\code_{+} \to \code \mid \stepNum_{+}],
        \\
        p_{-} := \Pr[\code \to \code_{-} \mid \stepNum_{-}].
    \end{gather*}
    Then, for all $\code_{+} \in \setPSimp$ and $\code_{-} \in \setNSimp$,
    \[
        \Pr[\code_{+} \to \code_{-} \mid \stepNum_{+} + \stepNum_{-}]
        \geq p_{+} p_{-} > 0.
    \]
    Hence every $\code_{-} \in \setNSimp$ is reachable from every $\code_{+} \in \setPSimp$, so $\setNSimp \subseteq \setPSimp$. By symmetry, we also obtain $\setPSimp \subseteq \setNSimp$. Therefore,
    \[
        \setPSimp = \setNSimp.
    \]
\end{proof}
}

\begin{definition}[Rule-based Partition]
\label{apx:def:partition}
    Following \cref{apx:tho:partition}, given a high-quality code space $\hqSpace$ and an ergodicity rule set $\ergRuleSet$, the induced rule-based partition $\partition$ is
    \[
        \partition = \curly{\funcErgEqSpace_{1}, \funcErgEqSpace_{2}, \dots, \funcErgEqSpace_{n}}
    \]
    and it satisfies the following properties (abbreviating $\partition$ as $\partitionSimple$):
    \[
    \begin{aligned}
        &\forall \eqSpace \in \partitionSimple, \eqSpace \subseteq \hqSpace; \\
        &\bigcup_{\eqSpace \in \partitionSimple}\eqSpace = \hqSpace; \\
        &\forall \eqSpace_{+},\eqSpace_{-} \in \partitionSimple, \eqSpace_{+} \cap \eqSpace_{-} = \varnothing.
    \end{aligned}
    \]
    Thus, $\partition$ is precisely the partition of $\hqSpace$ induced by $\ergRuleSet$.
\end{definition}

\section{Impossibility Results}
\label{apx:sec:impossibility}

\subsection{Assumption and Robustness Impossibility}

In this section, we will describe our only assumption, which is intuitive and has been supported by experiments.

First, we define the target of our assumption, distribution consistency.

\begin{definition}[Distribution Consistency]
\newcommand{\anyDist}{\mathcal{Q}}
\label{apx:def:consistency}
    We define $\anyDist(\codeSpace)$ to denote a distribution over the code space $\codeSpace$, where $\anyDist$ is independent of the watermarking scheme.
    Given a partition $\partition$, a watermarking scheme $\wmScheme=(\watermark,\detect)$ satisfies the \textit{distribution consistency} property if, for all $\anyDist$, $\prompt \in \promptSpace$, $\code \in \codeSpace$, and $\secretKey \in \keySpace$, given $\code \sim \anyDist(\codeSpace), \detect_{\secretKey}(\prompt, \code) \sim \unknownDist$, we have:
    \[\begin{array}{c}
        \text{if } \code \sim \anyDist(\eqSpace),
        \text{then } \detect_{\secretKey}(\prompt, \code) \sim \unknownDist.
    \end{array}\]
    That is, the distribution of $\detect_{\secretKey}$ within each equivalent space $\eqSpace \in \partition$ remains the same as the distribution observed over the entire code space.
\end{definition}

Recall that distribution consistency is a necessary and sufficient condition for the independence between $\ergRuleSet$  and $\detect$.
Moreover, the $\detect(\prompt, \code)$ among the whole code space $\codeSpace$ follows the Bernoulli distribution, $\bern(\fpr)$.

\begin{assumption}[Impossibility Assumption]
\label{apx:asm:main}
    With implementable code transformation rules, distribution consistency (\cref{apx:def:consistency}) can be achieved for N-gram-based watermarking schemes.
\end{assumption}

\textbf{Discussion of \cref{apx:asm:main}}: 
First, distribution consistency is clearly not guaranteed for arbitrary rule sets.
For example, if a rule set only makes minor surface changes, many literal features may be preserved, and the resulting samples can remain biased under N-gram-based watermark detectors.
An extreme case is a rule set containing only the identity rule, which produces no perturbation at all.

Nevertheless, the assumption becomes more plausible for sufficiently expressive transformation rule sets.
Modern syntax- and semantics-preserving transformations can substantially vary the N-gram features within an equivalent space.
For example, variable names, string or numeric constants, local control/data flow, and expressions can all be perturbed to different degrees, often independently of the N-gram-level features used by watermark detectors.
By contrast, the components that tend to remain stable under such transformations are usually semantic in nature, such as design patterns, high-level control/data flow, and functionality.
Since N-gram-based watermarking schemes are not aware of these semantic features, it is reasonable to assume that their detectable N-gram features are largely independent of the equivalent-space partition.


We can think of distribution consistency as the product of a game between watermarking schemes and attackers. 
If the watermarking scheme can watermark and detect code properties at a higher level, e.g., syntax or even functional level, while the rule set can only union codes with low-level equivalence, the distribution inside rule-based equivalent space will be seriously biased. For example, if the watermarking scheme can watermark and detect the AST pattern, while the rule set can only disturb variable names, the $\detect$  will return the same results inside each equivalent space.
On the other side, if the watermarking scheme can only watermark and detect code properties at a low level, like the N-gram level, the distribution consistency is easier to satisfy by existing syntax or semantic level transformation rules.

Based on the discussion above, we make \cref{apx:asm:main} and empirically test it in \cref{sec:dist_test_ext}.

\begin{theorem}[Impossibility Theorem]
\label{apx:tho:main}
    With \cref{asm:main}  satisfied, given large enough $\stepNum$, there exists an ergodicity rule set $\ergRuleSet$  that for all N-gram-based schemes with false positive rate $\fpr$, the $\walk(\code \mid \code \in \codeSpace_{\secretKey}, \ergRuleSet, \stepNum)$  algorithm can perform as an attacker and $(1-\fpr)$-break the watermarking.
\end{theorem}

\begin{proof}[Proof of \cref{apx:tho:main}]
    We remind that:
    \begin{itemize}
        \item The secret key $\secretKey$  is sampled randomly and independently of $(\prompt, \code) \in \promptSpace \times \codeSpace$, which ensures that we can bound the false positive rate without needing to make assumptions on the unknown human-generated data distribution.
        \item For each ergodicity rule set, given long enough $\stepNum$, start from a watermarked code $\code$, after $\walk(\code, \ergRuleSet, \stepNum)$  runs, the results (and the stationary distribution $\statDist$) only rely on: 
        \textbf{1)} Which equivalent spaces $\eqSpace^{\ergRuleSet}_{\code}$  the original $\code$  belongs to. 
        \textbf{2)} The stationary distribution $\statDist$  of $\graph^{\ergRuleSet}_{\code}$.
        \item We assume the distribution consistency (\cref{asm:main}), i.e., the independence between $\ergRuleSet$  and $\detect$.
    \end{itemize}
    Therefore, if we set: 
    \begin{itemize}
        \item $\adversary(\prompt, \code) := \walk(\code, \ergRuleSet, \stepNum)$  (the prompt $\prompt$  is provided as an input here, but note that the algorithm $\walk$  does not rely on it),
        \item For $\partition = \curly{\eqSpace_1, \eqSpace_2, \dots, \eqSpace_m}$, the probability distribution $\spaceDist \in \real^m$  has $\spaceDist_j = \expect[\code \in \eqSpace_j \mid \code \in \codeSpace_{\secretKey}],$
    \end{itemize}
    we have:
    \[\begin{aligned}
        & \expect(\indicator[\detect_{\secretKey}(\prompt, \adversary(\prompt, \code)) = 0] \mid \code \in \codeSpace_{\secretKey}) \\
        = & 1 - \sum_{\eqSpace_j \in \partition} (
            \spaceDist_j \cdot \expect [
                \detect_{\secretKey}(\prompt, \code_i) \mid \code_i \in \eqSpace_j
            ] 
        ) \\
        = & 1 - \sum_{\eqSpace_j \in \partition} (
            \spaceDist_j \cdot
            \sum_{\code_i \in \eqSpace_j} (
                \statDist_i \cdot \expect [
                    \detect_{\secretKey}(\prompt, \code_i)
                ] 
            )
        ) \\
        = & 1 - \sum_{\eqSpace_j \in \partition} (
            \spaceDist_j \cdot
            \sum_{\code_i \in \eqSpace_j} (
                \statDist_i \cdot \fpr
            )
        )
        = 1 - \fpr
    \end{aligned}\]
\end{proof}

In other words, when distribution consistency is satisfied, the distribution of the watermarking detectable N-gram feature among each equivalent space does not differ from the distribution among the whole code space $\codeSpace$.
Therefore, after the obfuscation, the watermark detector cannot distinguish the watermarked code from benign code samples selected from the human-written space, making the "afterward FNR" increase to a very high level of $1 - \fpr$, effectively nullifying watermark detection.

The above impossibility theorem relies on the condition \textit{large enough $\stepNum$}. To better understand the efficiency of our attack, with the mixing time $\stepNum_{i}(\aRate)$  defined in Appendix \ref{apx:sec:chain_modeling}, we further prove that:

\begin{theorem}[Lower Bound of Attacked FNR]
\label{apx:tho:lower_bound}
    With \cref{apx:asm:main}  satisfied, given $\stepNum \geq \stepNum_{s}(\aRate)$, in which $\code_s$  is the original watermarked code and also the starting point of the random walk, there exists an ergodicity rule set $\ergRuleSet$  that for all N-gram-based schemes with a false positive rate $\fpr$, the $\walk(\code_s \mid \code_s \in \codeSpace_{\secretKey}, \ergRuleSet, \stepNum)$  algorithm can perform as an attacker and $(1-\aRate-\fpr)$-break  the watermarking.
\end{theorem}

\begin{proof}[Proof of \cref{apx:tho:lower_bound}]
\newcommand{\abbDetect}{\textsf{D}}
    Given $\stepNum \geq \stepNum_{s}(\aRate)$  achieved, in the worst case, the whole probability distribution is biased toward $\detect_{\secretKey}(,)=1$. Setting $\Delta_i := |\distVec(,)_i-\statDist_i|$  and $\abbDetect_i := \detect_{\secretKey}(\prompt, \code_i)$, we have:
    \[\begin{aligned}
        & \sum_{\code_i \in \eqSpace}  ( 
            \distVec(\code_s, \stepNum)_i \cdot \abbDetect_i
        ) 
        =  \sum_{
            \substack{\code_i \in \eqSpace \land \\ \abbDetect_i = 1}
        } \distVec(\code_s, \stepNum)_i \\
        \leq & \sum_{
            \substack{\code_i \in \eqSpace \land \\ \abbDetect_i = 1}
        } [\statDist_i + \Delta_i] 
        = \sum_{
            \substack{\code_i \in \eqSpace \land \\ \abbDetect_i = 1}} \Delta_i + 
            \sum_{\code_i \in \eqSpace}[\statDist_i \cdot \abbDetect_i].
    \end{aligned}\]
    In the worst case, the bias increases the probabilities of the $\abbDetect_i=1$ cases, with a symmetric reduction on $\abbDetect_i=0$ cases. Therefore:
    \[
        \sum_{
            \substack{\code_i \in \eqSpace \land \\ \abbDetect_i = 1}} \Delta_i
        \leq \frac{1}{2} \sum_{\code_i \in \eqSpace}\Delta_i
        = \tvDist{\distVec(,)}{\statDist}
        \leq \aRate.
    \]
    Substituting into the previous calculation, with $\stepNum \geq \stepNum_{s}(\aRate)$, we get:
    \[\begin{aligned}
        & \expect(\indicator[\detect_{\secretKey}(\prompt, \adversary(\prompt, \code_s)) = 0] \mid \code_s \in \codeSpace_{\secretKey}) \\
        = & 1 - \sum_{\eqSpace_j \in \partition} (
            \spaceDist_j \cdot \expect [
                \sum_{\code_i \in \eqSpace_j}  ( 
                    \distVec(\code_s, \stepNum)_i \cdot \abbDetect_i
                )
            ] 
        ) \\
        \geq & 1 - \sum_{\eqSpace_j \in \partition} (
            \spaceDist_j \cdot \expect [
                \aRate + 
                \sum_{\code_i \in \eqSpace_j}  ( 
                    \statDist_i \cdot \abbDetect_i
                )
            ] 
        ) \\
        = & 1 - \aRate - \fpr
    \end{aligned}\]
\end{proof}


Note that as "afterward FNR," the $1 - \aRate - \fpr$  is still too high to be an acceptable value, for the time complexity calculation shows in \cref{tho:mixing_cal}  that $\stepNum_{s}(\aRate)$  has a linear dependence on $\ln \aRate^{-1}$, meaning that $\aRate$  can be set to a small enough value with a slight affection on the $\stepNum_{s}(\aRate)$.

\section{Cost Analysis of Attacks}
\label{sec:cost_analysis}
{
\newcommand{\bigO}[1]{\mathit{O} \left( #1 \right)}
\newcommand{\order}[1]{\Theta \left( #1 \right)}
\newcommand{\spaceSize}{\mathit{N}}
\newcommand{\codeLen}{\mathit{L}}
\newcommand{\segLen}{\mathit{l}}
\newcommand{\outDeg}{\mathit{d}}
\newcommand{\receNum}{\mathit{\wp}}

\newcommand{\constA}{\alpha}
\newcommand{\constB}{\beta}
\newcommand{\constC}{\partial}

\newcommand{\canonPath}{\zeta}
\newcommand{\pathSet}{\Upsilon}
\newcommand{\congestion}{\varrho}
\newcommand{\dof}{d^{\mathsf{free}}}

\subsection{Overall Analysis}
\label{sec:cost_overall}

In other sections, we prove that when $\stepNum$ exceeds the mixing time and distribution consistency is satisfied, the robustness of the watermarking scheme can be effectively broken.

In this section, we estimate the order of this \textit{sufficiently large} $\stepNum$  and demonstrate that an attack tends to incur significantly lower time and space costs compared to the cost of code generation.

We emphasize that the time and space complexity should be various for different designs of the rule set. Therefore, we add intuitive and implementable additional definitions to simplify the calculation. The following analysis may not apply universally. Instead, our result should be considered an approximate solution applicable to general scenarios.

Our attack proceeds by first dividing the watermarked code into segments or code blocks. These segments only need to be sufficiently long to satisfy distribution consistency. Based on our experiments, this required length is relatively small. In fact, all three watermarking schemes lost the detection ability consistently during our experiment, and the $90$-th percentile of segment lengths was $275.0$, which can serve as a reference value for setting $l$.
Next, for each sub-segment, we perform iterative random walks until the mixing time is reached.





For the following proof, we set several symbols:

\begin{itemize}
    \item Given $\eqSpace \in \partition$, let the space size $\spaceSize = \sizeof{\eqSpace}$.
    \item Let the length of the watermarked code be $\codeLen$  and the length of the split code segment be $\segLen$.
    \item Let the receptor number of the code segment be $\receNum$. As for the concept \textit{receptor}, it represents the code component that is targeted by rules. E.g., code variables for variable modification rule, comments for comment modification rule, etc.
    \item Let the out-degree of each vertex in the code transformation graph be $\outDeg$.
\end{itemize}

We can follow the general assessment that the complexity of the action derivation algorithm $\eqRule$  is $\bigO{\segLen}$  and $\trans$  is $\bigO{1}$, for both time complexity and space complexity, since $\eqRule$  needs to iterate the code and find the applicable actions, and each action performed by $\trans$  only leads to modifications of constant order. Moreover, since the new modifications after each call of $\trans$  are of constant order and the $\eqRule$  in the next round only needs to analyze the updates, we can assess the complexity of $\eqRule$  as $\bigO{1}$  if not the initial call of $\eqRule$.

Therefore, $\bigO{\segLen}$  is the overall space complexity, and $\bigO{\frac{\codeLen}{\segLen} \cdot (\segLen + \stepNum)}$  is the time complexity. The key point is to assess the mixing time $\stepNum$.

In the following sections (\cref{tho:mixing_cal}), we prove that $\stepNum_{i}(\aRate) \leq \bigO{\segLen^2 \ln \aRate^{-1} + \segLen^3}$. In the case that $\codeLen$  is relatively short and we do not split the code into segments ($\codeLen = \segLen$), the overall time complexity of our attack will be $\bigO{\codeLen^2 \ln \aRate^{-1} + \codeLen^3}$. When we consider the situation that $\codeLen \gg \segLen$  and we apply the split, the overall complexity will be $\bigO{\codeLen \cdot (\segLen \cdot \ln \aRate^{-1} + \segLen^2)} = \bigO{\codeLen}$, since $\codeLen \gg \segLen$  (as shown in our experiments, $\segLen$ does not need to be large).

Compared with code generation, nowadays, transformer-based generative models can be considered as having time and space complexity $\bigO{\codeLen^2}$  with a well-known high constant factor. Since our attack generally only needs lightweight CPU computation and memory storage, our algorithm with $\bigO{\segLen}$  space and $\bigO{\codeLen}$  time needed requires much less computation than code generation. 
Even if we do not split the watermarked code during the attack and obtain $\bigO{\codeLen^2 \ln \aRate^{-1} + \codeLen^3}$  time complexity, the algorithm still tends to be faster than LLM generation when we have small scale $\codeLen$  due to generative model's high constant factor. 


\subsection{Congestion and Canonical Paths}
\label{sec:congestion}
To assess the mixing time, we introduce classic and effective tools from Markov chain theory, \textit{congestion}  and \textit{canonical paths} \cite{jerrum2003counting}.
Given code transformation graph $\graph^{\ergRuleSet}_{\code} = (\vertexSet, \edgeSet, \weightSet)$, vertex indices $i,j,I,J \in \nature$, we have:
\begin{itemize}
    \subdef{Canonical Path}
        for any pair $\vertex_i, \vertex_j \in \vertexSet$, define a canonical path $\canonPath_{ij} = (\vertex_i = \dotsc = \vertex_j)$  from $\vertex_i$  to $\vertex_j$  through directed edges, and let $\pathSet := \curly{\canonPath_{ij} \mid \vertex_i, \vertex_j \in \vertexSet}$  be the set of all canonical paths.
    \subdef{Congestion}
        The \textit{congestion}  $\congestion$  of the graph is defined by:
        \[
            \congestion(\pathSet) = \max_{\edge_{IJ} \in \edgeSet} \curly{\frac{\sum_{i,j: \canonPath_{ij} \text{ uses } \edge_{IJ}} \statDist_i \statDist_j \sizeof{\canonPath_{ij}}}{\statDist_I \transMat_{IJ}}} 
        \]
    \subdef{Congestion and Mixing Time}
        Denoting $\congestion := \congestion(\pathSet)$, the mixing time $\stepNum_i(\aRate)$  is bounded by:
        \[
            \stepNum_{i}(\aRate) \leq 2 \congestion(2 \ln \aRate^{-1} + \ln \statDist_i^{-1}).
        \]
\end{itemize}

Canonical paths and congestion provide us with an elegant approach to bound mixing times, in which we are allowed to flexibly define the form of our canonical paths.

\subsection{Graph Definitions}

In this section, we begin the mixing-time calculation.
We define the concept \textit{receptor}, which means the element that receives the code transformation, for example, variables for the rule "Randomly rename variables" and dead code block for "Delete a dead code snippet" in \cref{apx:tab:erg_set_example}.

Given the notion that the size of $\eqSpace \in \partition$  is $\spaceSize$  and the out-degree of transformation graph is $\outDeg$, we stress the following additional definitions, which are intuitive and implementable, to simplify the calculation.

We clarify that the following mixing time estimation is derived under these specific assumptions. While it is exact in such constrained settings, it can also serve as an approximate solution in more general scenarios.

\paragraph{Code Length}
When random walk on code segments inside the same transformation graph, all lengths of code are of the order $\order{\segLen}$, i.e., the ratios among them are of the constant order. It is less reasonable if we obtain a result after a random walk that is too short or too long compared with the original watermarked code, and we also can constrain the range of code length during the random walk by the implementation of our rule set.

\paragraph{Independent Receptors}
All receptors in a code segment are satisfied: 
\textbf{1)} Non-overlapped. One element of code can only receive one rule. For example, in the rule set of \cref{apx:tab:erg_set_example}, the variable names inside a dead code snippet can be modified by either the variable renaming rule or the dead code deletion rule, but not both.
\textbf{2)} Pre-defined. At the initial stage, the rule algorithms will identify receptors with fixed positions. For example, the comment-adding rule will identify positions that can accept a new comment as its receptors, together with existing comments positions as candidates. At the following stages, it will only select a position without comments from these pre-defined positions for adding.
\textbf{3)} State-changeable. Each receptor can be modified to have multiple states but will not disappear during the random walk. For example, after a transformation deletes a comment, the original position turns into a receptor of the comment addition rule.


Therefore, the receptor number $\receNum = \order{\constB \cdot \segLen} = \order{\segLen}$, in which $\beta \leq 1$  is a constant and denotes the expected number of receptors distributed per code token.

\paragraph{Out-degree}
The ratio between the out-degree values of two vertices in the same transformation graph is of the constant order. Because for different code segments with similar lengths, with the same rule set, the number of possible transformation actions should be of the same order. Therefore, we can denote out-degree $\outDeg = \order{\constA \cdot \receNum} = \order{\segLen}$, in which $\constA \geq 1$  is a constant and denotes the expectation number of actions that rule set $\ergRuleSet$  can perform on each receptor location.

\subsection{Mixing Time Calculation}

In this section, we will discuss subexpressions separately and conduct the mixing time from congestion.

\begin{definition}[Canonical Path for Code Transformation Graph]
\label{def:code_trans_canon}
    For any pair $\vertex_i, \vertex_j \in \vertexSet$, a canonical path from $\vertex_i$  to $\vertex_j$  satisfied:
    \begin{itemize}
        \subdef{Shortest Path} 
            The $\canonPath_{ij}$  can not be shorter than other paths from $\vertex_i$  to $\vertex_j$, formally: 
            \[
                \sizeof{\canonPath_{ij}} = \min \curly{ \sizeof{\canonPath} \mid \canonPath = (\vertex_i = \dotsc = \vertex_j)}.
            \]
            Therefore, we have $\sizeof{\canonPath_{ij}} = \order{\receNum} = \order{\segLen}$  since each receptor can be modified into the final stage within one shot.
        \subdef{In Sequence Modification}
            For edges in each $\canonPath_{ij}$, the corresponding receptors follow some queuing convention. 
            The receptor's rank in the queue is monotonically increasing along the canonical path.
    \end{itemize}
\end{definition}

Note that the \textit{in sequence modification}  definition is implementable. For example, the receptors are sorted by their hunk positions from front to back, and then we can execute the modifications for sequenced receptors one by one.

\begin{theorem}[Transition Matrix Elements are of $\order{\frac{1}{\outDeg}}$]
\label{tho:order_trans_mat}
    For all $\edge_{ij} \in \edgeSet$, the corresponding $\transMat_{ij} = \order{\frac{1}{\outDeg}}$.
\end{theorem}

\begin{proof}[Proof of \cref{tho:order_trans_mat}]
    From \cref{apx:def:random_transformation}, which defines an uniform and random algorithm to pick the next vertex during random walk, given out-degrees are of the order $\order{\outDeg}$, it follows immediately that for all $\edge_{ij} \in \edgeSet$, $\transMat_{ij} = \order{\frac{1}{\outDeg}}$.
\end{proof}

\begin{theorem}[Stationary Distribution Elements are of $\order{\frac{1}{\spaceSize}}$]
\label{tho:order_stat_dist}
    For stationary distribution $\statDist$, it satisfied:
    \[
        \forall i \in \curly{1,2,\dots,\spaceSize}, \statDist_i = \order{\frac{1}{\spaceSize}}
    \]
\end{theorem}

\begin{proof}[Proof of \cref{tho:order_stat_dist}]
    First, from \textit{inverse rule} property defined in \cref{def:erg_rule_set}, for each vertex in the $\graph^{\ergRuleSet}_{\code}$, its out-degree is equal to in-degree.
    Then, we assume after $\stepNum$  steps of random walk:
    \[
        \forall \vertex_i \in \vertexSet, \distVec(\code, \stepNum)_i = \frac{\outDeg_i}{\sum_{\vertex_k \in \vertexSet} \outDeg_k},
    \]
    in which $\outDeg_i$ denotes the out-degree of $\vertex_i$. The denominator in this equation means the sum of all vertices' out-degrees.
    Considering further step $\stepNum + 1$, for all $\vertex_i \in \vertexSet$, we have:
    \[\begin{aligned}
        & \distVec(\code, \stepNum + 1)_i = \sum_{\edge_{ji} \in \edgeSet} \distVec(\code, \stepNum)_j \weight_{ji} \\
        = & \sum_{\edge_{ji} \in \edgeSet} 
        \frac{\outDeg_j}{\sum_{\vertex_k \in \vertexSet} \outDeg_k} \cdot \frac{1}{\outDeg_j} = \outDeg_i \cdot \frac{1}{\sum_{\vertex_k \in \vertexSet} \outDeg_k} \\
        = & \distVec(\code, \stepNum)_i.
    \end{aligned}\]
    We can notice that after the assumption is satisfied, the random walk falls into its stationary. Due to the uniqueness of the stationary distribution, we can confirm that for all $\vertex_i \in \vertexSet$:
    \[
        \statDist_i = \frac{\outDeg_i}{\sum_{\vertex_k \in \vertexSet} \outDeg_k} = \order{\frac{\outDeg}{\sizeof{\vertexSet} \cdot \outDeg}} = \order{\frac{1}{\spaceSize}}.
    \]
\end{proof}

\begin{theorem}[Canonical Path Number Uses Particular Edge are of $\bigO{\spaceSize}$]
\label{tho:order_path_number}
    Given the $\canonPath$  definition in \cref{def:code_trans_canon}, for each $\edge_{IJ} \in \edgeSet$, we have:
    \[
        \sizeof{\curly{\canonPath_{ij} \mid \canonPath_{ij} \text{ uses } \edge_{IJ}}} = \bigO{\spaceSize}
    \]
\end{theorem}

\begin{proof}[Proof of \cref{tho:order_path_number}]
    Given definition in \cref{def:code_trans_canon}, we further define the degrees of freedom on receptors ranked $1$  to $\receNum$  are $\dof_1, \dots, \dof_{\receNum}$. Therefore, the space size $\spaceSize = \prod_{k=1}^{\receNum} \dof_k$.
    
    For each path uses $\edge_{IJ}$, if the corresponding receptor of $\edge_{IJ}$  is ranked $K$. We can notify that the sub-path before $\edge_{IJ}$  is limited to only involve the modifications on receptors ranked $1$  to $K$, and vice versa, the sub-path after $\edge_{IJ}$  can only involve receptors $K$-th  to $\receNum$-th. 

    Therefore, the degree of free of $\vertex_i$, i.e., the path's starting vertex, is not bigger than $\prod_{k=1}^{K} \dof_k$, and for $\vertex_j$  is not bigger than $\prod_{k=K}^{\receNum} \dof_k$. We consider all combinations of starting vertices and ending vertices to get the following:
    \begin{gather*}
        \sizeof{\curly{\canonPath_{ij} \mid \canonPath_{ij} \text{ uses } \edge_{IJ}}} 
        \\
        \leq \prod_{k=1}^{K} \dof_k \cdot \prod_{k=K}^{\receNum} \dof_k = \bigO{\spaceSize}.
    \end{gather*}
\end{proof}

\begin{theorem}[Mixing Time Calculation]
\label{tho:mixing_cal}
    Given initial code segment $\code$, segment length is of $\order{\segLen}$, the mixing time $\stepNum_{i}(\aRate)$  of random walk satisified: $\stepNum_{i}(\aRate) \leq \bigO{\segLen^2 \ln \aRate^{-1} + \segLen^3}.$
\end{theorem}

\begin{proof}[Proof of \cref{tho:mixing_cal}]
    Given \cref{tho:order_path_number}, we have:
    \[
        \ln{\spaceSize} \leq \ln{\left[\max\curly{\dof_1, \dots, \dof_{\receNum}}^{\receNum}\right]} = \bigO{\segLen}.
    \]
    Together with \cref{def:code_trans_canon}, \cref{tho:order_trans_mat}, and \cref{tho:order_stat_dist}, we have:
    \[\begin{aligned}
        \stepNum_{i}(\aRate) \leq & 2 \congestion(2 \ln \aRate^{-1} + \ln \statDist_i^{-1}) \\
        = & \bigO{\frac{\spaceSize \cdot \frac{1}{\spaceSize} \cdot \frac{1}{\spaceSize} \cdot \constC \receNum}{\frac{1}{\spaceSize} \cdot \frac{1}{\outDeg}} \cdot (2 \ln \aRate^{-1} + \ln\spaceSize)} \\
        = & \bigO{\segLen^2 \cdot (2 \ln \aRate^{-1} + \ln\spaceSize)} \\
        = & \bigO{\segLen^2 \ln \aRate^{-1} + \segLen^3},
    \end{aligned}\]
    which means following our code transformation model, a rapid mixing \cite{guruswami2000rapidly}  would be achieved. Together with our analysis in \cref{sec:cost_overall}, this time complexity leads to a significantly lower runtime cost than LLM-based code generation.
\end{proof}

}

\section{Impossibility of Impossibility}
\label{sec:imp_of_imp}

In this section, we demonstrate the unfeasibility of the previous work's impossibility proof \cite{zhang2023watermarks}  on program generation tasks, in which the authors made two critical assumptions:

\begin{itemize}
\item \textbf{Quality oracle}: The attacker has access to a \textit{quality oracle}, allowing efficient verification of whether a disturbed result still meets the original prompt's requirements, i.e., belonging to the high-quality space.
\item \textbf{Perturbation oracle}: The attacker has a \textit{perturbation oracle}, which can efficiently generate random disturbances to responses while maintaining their quality. The perturbation oracle relies on a quality oracle to validate whether a randomly perturbed answer still belongs to the high-quality space.
\end{itemize}

Under these assumptions, the authors proved that no watermarking scheme could retain its robustness. They argued that these assumptions are practically reasonable because of the heuristic notion that "verification is easier than generation." However, we argue that in programming generation tasks, these assumptions, especially the existence of a quality oracle, do not hold either theoretically or practically.

\paragraph{Theoretical Impossibility}
Specifically, unlike natural language scenarios where verification might indeed be simpler than generation, verifying correctness in programming generation needs to observe the running behavior and is fundamentally challenging. The quality oracle described would need to check whether the generated code precisely matches a prompt describing the required program behavior. However, this universal verification task is theoretically impossible due to fundamental limitations rooted in computability theory.
It is deeply connected to the well-known \textit{halting problem}  \cite{kleene1952introduction}. The halting problem demonstrates that no algorithm can universally determine whether an arbitrary program halts (terminates) or continues indefinitely for all possible inputs.

Building upon the halting problem, Rice's theorem \cite{rice1953classes}  generalizes this concept further, asserting that virtually all meaningful (non-trivial) behavioral properties of programs, such as correctness, termination, or compliance with certain specifications, are undecidable. In simpler terms, Rice's theorem implies that there is no universal verifier capable of reliably assessing program behavior across all possible cases. This limitation is fundamental, not just an occasional exception.

Therefore, even for a seemingly straightforward prompt such as "Generate a program that halts," constructing an accurate quality oracle would involve solving the halting problem itself, which is known to be impossible. Extending this reasoning, we conclude that most practical programming prompts inherently involve verifying non-trivial semantic properties, making a universal quality oracle fundamentally unattainable.

For a more practical example, consider a prompt such as "generate a Python server demo." Implementing a quality oracle for this task would require determining whether a perturbed program is a valid Python server and verifying if its runtime performance (e.g., latency, throughput, correctness under various conditions) remains consistent with the original unperturbed program. This form of comprehensive runtime and semantic verification is also theoretically impossible, as it would necessitate solving undecidable problems.

\paragraph{Practical Impossibility}
Even if we add many restrictions (e.g., timeout, using LLMs to generate test cases, considering only codes that are runnable in $\adversary$'s sandbox, etc.) to make it theoretically implementable, achieving the quality oracle is still an open question and tends to consume large amounts of computational resources, leading to an attack that is too inefficient to be realistic.

Many works highlighted the difficulty of automating generating test cases from natural language business rules. Even though state-of-the-art generative models can be leveraged, there are still overwhelming challenges, including high intellectual demand, uncontrollable and intractable outputs, limited domain knowledge in pre-training models, test-oracle problems, rigorous evaluations, and high implementation costs in the real-world applications \cite{corriveau2014requirements, xue2024llm4fin, wang2024software}. 

We can also estimate the difficulty from past practice. An acceptable code evaluation is both hard and resource-consuming, even with ground truth code given. 
EvalPlus \cite{liu2023is}  needs a state-of-the-art model to generate around 30 seed inputs and generate 1000 additional inputs using a one-hour budget, for each task. 
LiveCodeBench \cite{jain2024livecodebench}  needs to generate other code segments as input generators, and there are 2 random and 4 adversarial input generators needed for each task. 
Considering the quality oracle, with only a natural-language description and reference code as input, it can only be harder without ground truth code to guide the test generation. CodeT \cite{chen2022codet}  needs to generate around 50 other candidate solutions (1000 for harder tasks) to perform cross-validation, and they generate 100 test cases for each task.
All works above are required to call state-of-the-art generative models.

Therefore, this is the opposite of the previous work's assumption that, even if we add restrictions to get around theoretical impossibilities, the code evaluation would be likely to consume similar or more computational resources than a generation, leading to a theoretically impossible and low cost-effectiveness attack.
This is also the motivation of our \textit{code transformation rule}  modeling.

In summary, to build a \textit{quality oracle}, there needs to be a theoretically impossible evaluator that tests whether codes fulfill requirements from the prompt. Even considering multiple restrictions are added to bypass the theory barrier, it still tends to require higher computational resources to evaluate the code than code generation.
We do not deny the feasibility of applying more advanced code perturbation methods and semantics checkers (e.g., use LLMs to re-generate code and use EvoSuite \cite{fraser2011evosuite}  to ensure equivalence where possible). However, any such attempt can only be seen as performing a random walk on a larger rule-based equivalent space but not the entire high-quality space, falling into the modeling in this paper, and an analysis of distribution consistency is needed.

\section{Discussion: Semantic Awareness of Code Watermarking}
\label{sec:sem_aware}

In this section, we will discuss a possible path to improving the robustness of code watermarking schemes, i.e., semantic awareness.

In \cref{sec:impossibility}, we analyze that if the randomized watermarking algorithms operate at the N-gram level without the awareness of syntax or semantics, the distribution consistency is more achievable since the transformation rules are generally syntax- or semantics-based.

On the other hand, in Appendix \ref{sec:imp_of_imp}, we also show that it is impossible to random walk or obtain semantic equivalence among the entire high-quality space $\hqSpace$, since they all rely on semantics comparison or semantics understanding, which would run into the undecidability barrier captured by Rice's theorem \cite{rice1953classes}, and leaves space for watermarking robustness.

Suppose we can introduce bias in the semantics of the generated code and detect the semantic features. In that case, attackers will need similar or higher computational costs to disturb the watermark than code generation.

As an example, one semantics-related component would be third-party API calling. Containing rich semantics, disruption on it needs an extensive understanding of semantics (e.g., transforming \texttt{np.square(matrix)}  to \texttt{matrix**2}  needs the knowledge of NumPy API). With obfuscators nowadays, third-party API calls generally remain even after obfuscation due to their semantics-related nature.

Consider a task that generates long code with multiple third-party API calls. If we have a watermarking scheme that can 1) apply hidden strategies on how to select and call third-party APIs and 2) detect these hidden patterns in the sequence of API calls, the robustness under attack would be kept to a considerable extent. The difficulty of attacking this watermarking scheme would be:

\begin{itemize}
    \item For attackers with a rule set that cannot disturb the API call, distribution consistency (\cref{asm:main}) would not hold since programs from each rule-based equivalent space share the same API calling feature, and the whole space can be considered watermarked.
    \item For attackers that are dedicated to disrupting the API-call features, a similar level of third-party API understanding compared with the code generation model is needed. Either dynamically disrupting API calls with a generative model or mining a comprehensive API transformation rule set would be costly and difficult to perform precisely. Meanwhile, a post-evaluation scheme (like the quality oracle in previous work \cite{zhang2023watermarks}) to bound the quality of transformed code would also be costly, as discussed in Appendix \ref{sec:imp_of_imp}.
\end{itemize}

Therefore, we consider a code watermarking scheme with high semantics awareness as robust.

\subsection{A Proof-of-Concept Design}
\label{sec:sem_aware_poc}

To make the path above concrete, we sketch a proof-of-concept scheme that operates over the sequence of third-party API calls in the generated program rather than over token N-grams.
\cref{alg:sem_poc} gives the pseudocode.

During generation, when the model is about to emit an API call, the scheme first identifies semantically valid candidate APIs for the current context at runtime.
The most recent $N-1$ API calls are then hashed to produce a seed that partitions these candidates into green and red sets, analogous to the token-level green list, and the model is biased toward a green-list choice; the first $N-1$ calls of a program are sampled without bias, following the same convention as the token-level scheme (Appendix \ref{sec:ngram_detail}).
For example, when both satisfy the required behavior, the model may prefer \texttt{pandas.isna(x)} over its documented alias \texttt{pandas.isnull(x)}.
During detection, the same keyed hash scores the observed API choices from the $N$-th call onward, and a z-score is computed from the number of scored calls assigned to the green set.

\begin{algorithm*}[t]
\caption{Proof-of-concept semantics-aware watermark over third-party API calls}
\label{alg:sem_poc}
    \KwIn{
        Model $\model$, prompt $\prompt$, key $\secretKey$,
        green fraction $\gamma$, logit bias $\delta$,
        context length $N$, detection threshold $\tau$
    }
    \textbf{Embedding:} \\
    Init generated code $\code \gets ()$, API-call history $A \gets ()$ \\
    \While{generation is not finished} {
        \eIf{$\model$ is about to emit a third-party API call} {
            \eIf{$\sizeof{A} \geq N-1$} {
                Init candidate set $C \gets$ semantically valid APIs for the current call site \\
                Init seed $s \gets H(\text{last } N-1 \text{ elements of } A, \secretKey)$ \\
                Partition $C$ by $s$ into a green set $G_s$ with fraction $\gamma$ and a red set $C \setminus G_s$ \\
                Add bias $\delta$ to the logits of APIs in $G_s$; sample the call $a$ \\
            } {
                Sample the call $a$ without bias \\
            }
            Append the tokens of $a$ to $\code$; append $a$ to $A$ \\
        } {
            Sample the next token from $\model$ and append it to $\code$ \\
        }
    }
    Return $\code$ \\
    \textbf{Detection} (given code $\code'$ and key $\secretKey$)\textbf{:} \\
    Init $A' \gets$ the sequence of third-party API calls extracted from $\code'$, green count $g \gets 0$, scored count $m \gets \sizeof{A'} - N + 1$ \\
    \For{$i \gets N$ \KwTo $\sizeof{A'}$} {
        Init candidate set $C \gets$ semantically valid APIs at the call site of $A'_i$ \\
        Init seed $s \gets H(A'_{i-N+1} \dots A'_{i-1}, \secretKey)$ \\
        Partition $C$ by $s$ into a green set $G_s$ with fraction $\gamma$ and a red set $C \setminus G_s$ \\
        \If{$A'_i \in G_s$} {
            Set $g \gets g + 1$ \\
        }
    }
    Return watermarked iff $z = \left(g - m\gamma\right) / \sqrt{m\gamma (1-\gamma)} > \tau$
\end{algorithm*}

This design survives the obfuscation attacks considered in this paper.
Obfuscators are generally syntax-based and do not perform semantic substitution between third-party APIs, so they preserve the selected API choices and their semantic call order.
Under such an obfuscator, every program in a rule-based equivalent space shares the same API-call feature, so the space is watermark-homogeneous with respect to the detector: distribution consistency (\cref{asm:main}) fails, and the detector retains discriminative power.
Substituting one third-party API for a semantically equivalent alternative would instead require API-level knowledge that existing rule sets lack, with the costs discussed above.

Note that showing the possibility of code watermarking robustness is not within the scope of this paper, and high implementation costs might accompany the path we discussed.

\begin{figure*}[t]
    \centering
    \includegraphics[width=\textwidth]{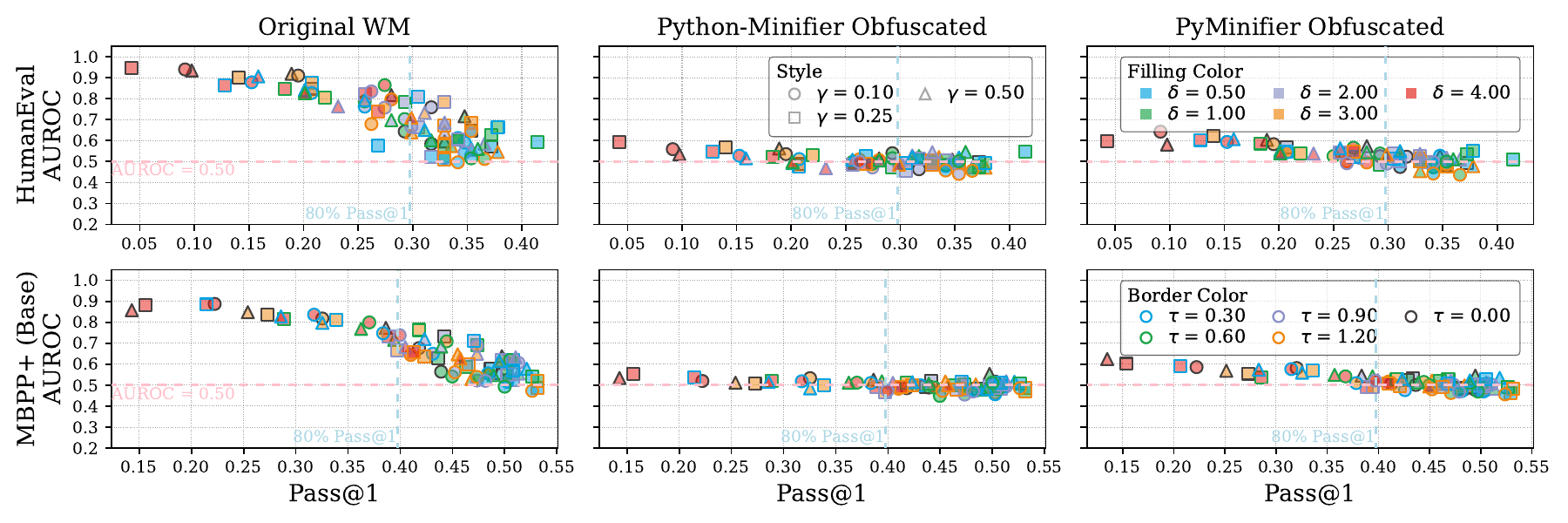}\\
    \vspace{0.0em}
    \includegraphics[width=\textwidth]{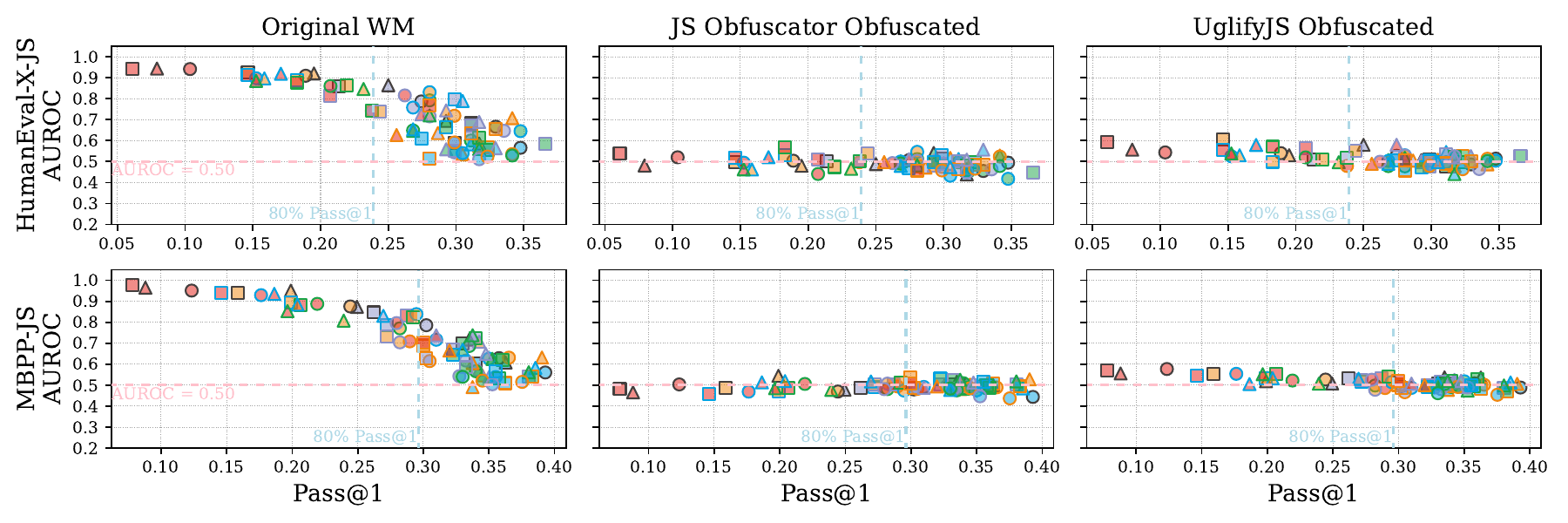}
    \caption{WLLM and SWEET \cite{kirchenbauer2023watermark, lee2023wrote} on DeepSeek-Coder-33B-Base \cite{guo2024deepseek}, before and after attack. The upper panel shows the effect of our attack on Python, and the lower panel shows the effect of our attack on JavaScript. The y-axis is AUROC and the x-axis is Pass@1; colors and markers denote hyperparameters. For LLaMA results, see \cref{fig:green_red_main}.}
    \label{fig:green_red_app}
\end{figure*}

\begin{figure}[htbp]
    \centering
    \includegraphics[width=0.95\linewidth]{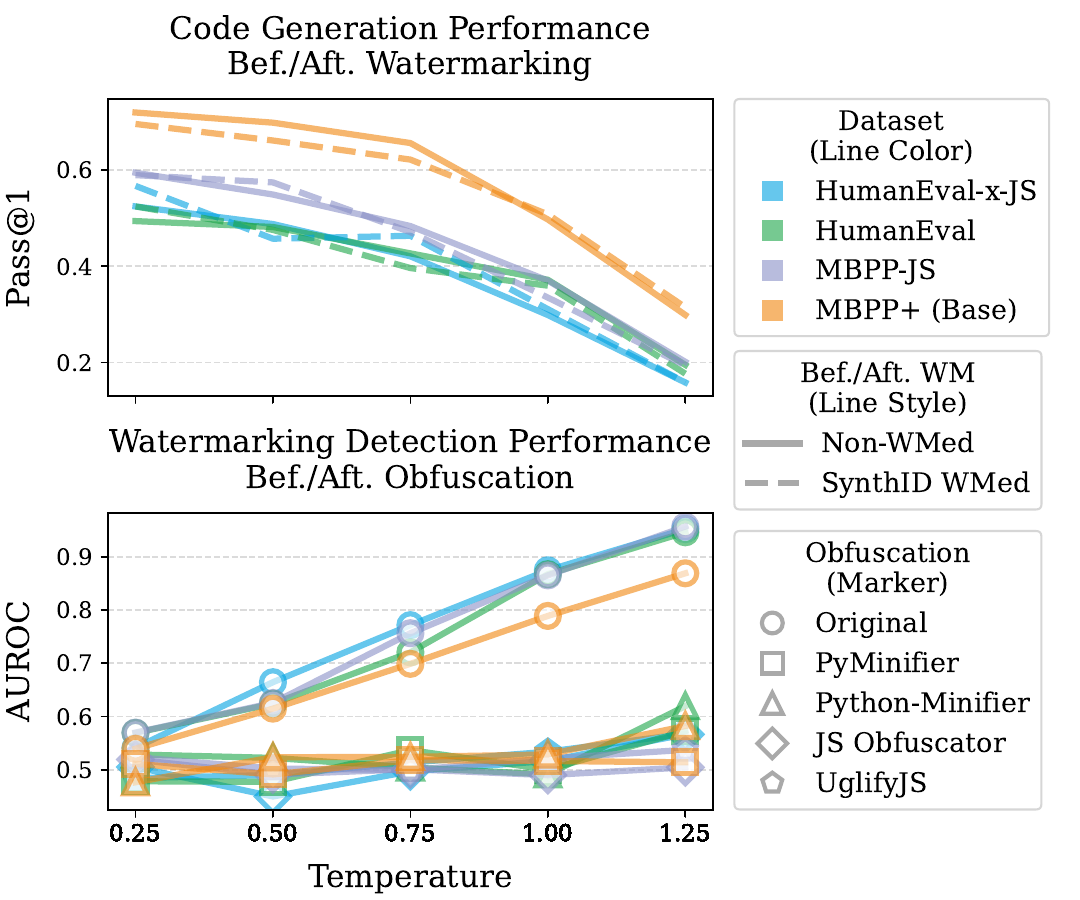}
    \caption{
    \underline{Upper}: 
    DeepSeek Coder Pass@1 changes with temperature. Comparison between SynthID \cite{dathathri2024scalable}  watermarked and non-watermarked code generation. 
    \underline{Lower}: 
    SynthID detection AUROC changes with temperature. Comparison between original SynthID watermarked code and obfuscated code.
    For results from the LLaMA 3.1, see \cref{fig:synthid_llama}.
    }\label{fig:synthid_deepseek}
\end{figure}

\section{Original Performances of Watermarking Schemes}
\label{src:original_performance}

Before studying robustness, we first analyze the original performance of the three selected watermarking approaches without an attack.

For distortionary watermarking schemes, i.e., SWEET and WLLM, the results on LLaMA against Python-Minifier and JS Obfuscator are shown in \cref{fig:green_red_main}. The DeepSeek results are shown in \cref{fig:green_red_app}.
The left-side sub-figures show that the original watermarking can achieve good detection performance. 
For LLaMA, all four code benchmarks have parameter combinations that lead to AUROC scores higher than $0.9$ while maintaining more than $80$ \%  of the same model's original non-watermarked performance. For example, under the WLLM setting $(\delta=4.00, \gamma=0.10)$, AUROC $= 0.97$  while maintaining $84$ \% non-watermarked performance.
For DeepSeek-Coder, all watermarking schemes also achieve AUROC $\simeq 0.8$ while maintaining $80$ \% non-watermarked performance.

The results for SynthID are shown in \cref{fig:synthid_llama}, as well as \cref{fig:synthid_deepseek} in the appendix. SynthID causes no obvious code-quality degradation for either model, due to its non-distortionary nature \cite{dathathri2024scalable}. 
For all experiments on SynthID, the average Pass@1 decrease is $-1.37 \times 10^{-3}$.
Considering detection performance under different settings, AUROC increases with temperature. 
The same AUROC trend can be observed from both models. With the temperature increasing from $0.25$  to $1.25$, among four benchmarks and two models, the average AUROC score increases from $0.56$  to $0.91$.

\textbf{Low-entropy Issue in Code Watermarking}:
From this result, we reaffirmed the low-entropy issue in code watermarking, which was also discussed in previous work \cite{lee2023wrote}.
To improve code quality, when generating code, the generation parameters are usually set to reduce the randomness, e.g., lower temperature or lower TopK. DeepSeek even directly applies greedy search when they evaluate on HumanEval-X \cite{guo2024deepseek}.
This exacerbates low-entropy characteristics in code generation and makes watermark detection harder, for the mainstream watermarking method requires higher entropy for better detection \cite{lee2023wrote, dathathri2024scalable}. In \cref{fig:synthid_llama}, we can observe this trend: as temperature decreases, code quality improves, but detection AUROC drops sharply. 
This provides additional evidence that N-gram-based watermarking is unsuitable for code content.

\section{Ideal Watermarking Scheme Under Attack}
\label{sec:exp_ideal_wm_attack}
\begin{table}
    \centering
    \small
    \begin{tabularx}{\columnwidth}{>{\raggedright\arraybackslash}X | c | c}
        \specialrule{1.2pt}{0pt}{0pt}
        & Pass@1 & AUROC \\ 
        \specialrule{0.9pt}{0pt}{0pt}
        Ideal Watermarking & 0.9634 & 0.9747 \\ \hline
        Attacked w/ UglifyJS & - & 0.5085  \\
        \specialrule{1.2pt}{0pt}{0pt}
    \end{tabularx}
    \caption{Ideal watermarking scheme performances before/after attack. The Pass@1 score after the attack is omitted because it is ensured to be identical with before.}
    \label{tab:ideal_result}
\end{table}

{
\begin{figure*}[ht]
    \centering
    \begin{minipage}[b]{0.49\textwidth}
        \centering
        \includegraphics[width=\textwidth]{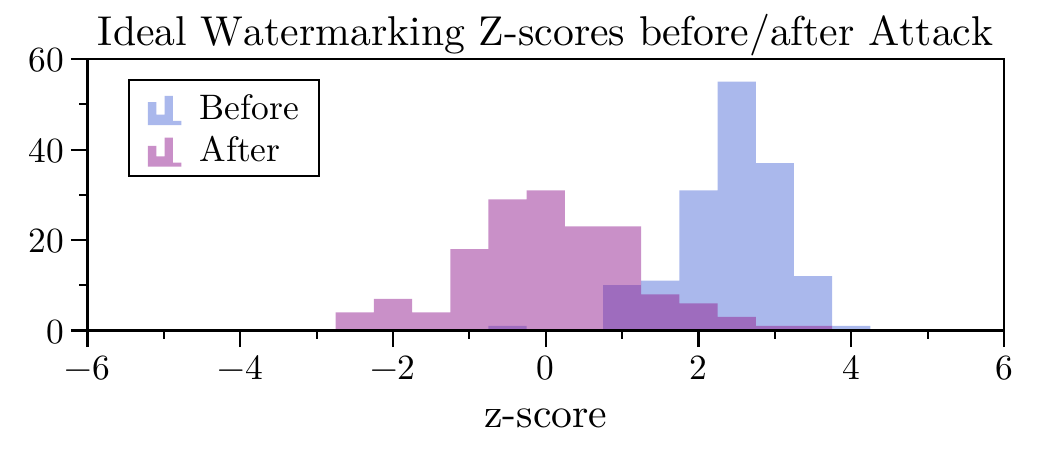}
    \end{minipage}
    \hfill
    \begin{minipage}[b]{0.49\textwidth}
        \centering
        \includegraphics[width=\textwidth]{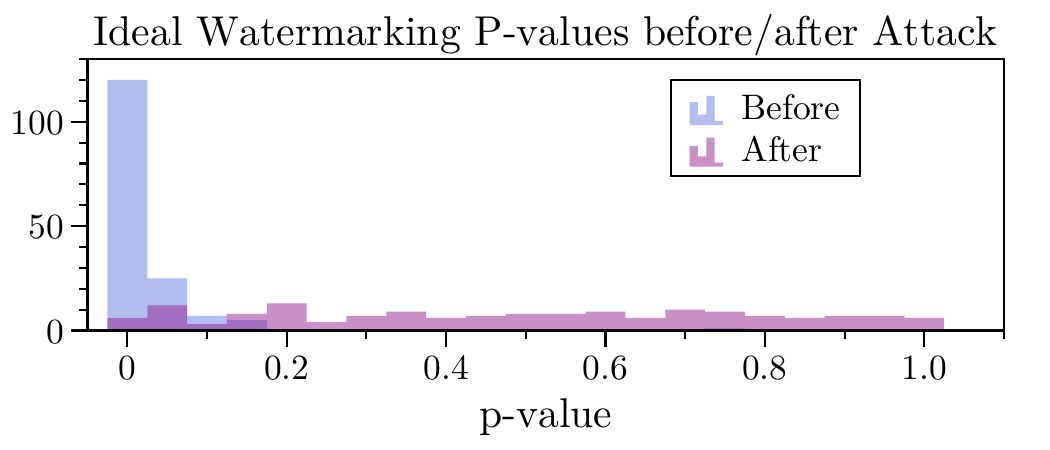}
    \end{minipage}
    \caption{
    \underline{Left}:
    Frequency distribution chart for z-scores of generated code from ideal watermarking scheme, before/after UglifyJS-based attack.
    \underline{Right}:
    Frequency distribution chart for p-values.
    }
    \label{fig:ideal-stat}
\end{figure*}
}

{
\newcommand{\isMarked}{\mathsf{isMarked}}
\newcommand{\tokenSpace}{\mathcal{V}}
\newcommand{\genTimes}{t_{\mathsf{gen}}}
\newcommand{\testSuiteBen}{\testSuite_{\mathsf{ben}}}
\newcommand{\finalR}{r_{\mathsf{final}}}
\newcommand{\wmR}{r_{\mathsf{wm}}}
\newcommand{\wmCode}{\code_{\mathsf{wm}}}

We construct an ideal watermarking scheme with unrealistically high detection capability and quality preservation, trying to answer a question: If we keep developing N-gram-based watermarking and its detection ability increases dramatically, would high robustness eventually emerge?

\Cref{alg:ideal_watermarking} shows the generation of our watermarking scheme. The $\isMarked$  is a hash algorithm that will return 1 when a \textit{marked} 5-gram is provided as input. We randomly selected half of the 5-grams as marked.
For each task, it repeatedly generates $\genTimes$  times (See \cref{line:rep_gen}; $\genTimes = 500$  in our experiment) with randomized generative model $\model$, which is a regular model without any watermarking related components.
For each generation, we use the test suite provided by the benchmark to filter out code that does not satisfy the prompt requirements (\cref{line:filter_w_bench}). 
Finally, if any of the generations can pass the test suite, we select the code with the highest ratio of marked 5-gram as the returned code (\cref{line:return_wm}).

For detection, we use the same $\isMarked()$  algorithm to calculate the ratio of marked 5-gram for each code under detection, then calculate corresponding $z$-score as the detection metric.

We claim that this watermarking algorithm has two properties that lead to unrealistically high detection capability and quality preservation, which realistic N-gram-based watermarking schemes are unlikely to achieve even in the future.
We aim to exploit the representational power of N-gram features as much as possible to combat our attacks.

\begin{algorithm}[!t]
\caption{Ideal Watermarking}
\label{alg:ideal_watermarking}
    {\small \tcp{$\tokenSpace$  is the space of tokens (vocabulary)} }
    {\small \tcp{$\isMarked: \tokenSpace^5 \to \curly{0,1}$  is the algorithm that can distinguish whether a 5-gram is marked, returning 1 denotes a marked 5-gram} }
    \KwIn{
        Generative Model: $\model: \promptSpace \to \codeSpace$,  \newline
        Prompt: $\prompt \in \promptSpace$, \newline
        Generation Times: $\genTimes \in \nature$, \newline
        Benchmark Given Test Suite: $\testSuiteBen$
    } \KwOut {
        Watermarked Code: $\wmCode \in \codeSpace$
    }
    Init Highest Watermarked Ratio $\finalR \gets 0$ \\
    Init Watermarked Code $\wmCode \in \codeSpace$ \\
    \For{each $i$  in $1,2, \dots, \genTimes$} { \label{line:rep_gen}
        Set $\code_i \gets \model(\prompt)$ \\
        \If {$\testSuiteBen(\prompt, \code_i) = 1$} { \label{line:filter_w_bench}
            Init Watermarked Ratio $\wmR \in \real$ \\
            Set $\wmR$  to the ratio of $\code_i$'s 5-grams which let $\isMarked()$  return 1 \\
            \If {$\finalR < \wmR$} {
                Set $\finalR \gets \wmR$ \\
                Set $\wmCode \gets \code_i$
            }
        }
    }
    \If {$\finalR = 0$} {
        Return No Result \\
    } \Else {
        Return $\wmCode$ \label{line:return_wm}
    }
\end{algorithm}

\begin{itemize}
    \item \textbf{Global Awareness}:
    Compared with realistic N-gram-based watermarking schemes that generally can only perform greedy algorithm per token, this scheme directly selects response with highest final global metric.
    \item \textbf{Test Oracle Assistance}:
    As we analyzed in Appendix \ref{sec:imp_of_imp}, high costs are needed for implementation of test suite (theoretically impossible for a generalized one). However, in this ideal scheme, we directly leverage the proprietary test suite in the benchmark to filter out unpassed generation, leading to an unrealistically high quality remaining.
\end{itemize}

Following \cref{sec:experiment_setting}  settings, we use HumanEval-X-JS \cite{zheng2023codegeex} and leverage UglifyJS-based attack on this ideal watermarking scheme.
The result is in \cref{tab:ideal_result}, and we gather the $z$-scores and $p$-values of the tasks, drawing the distributions in \cref{fig:ideal-stat}. 
We remind that our null hypothesis is $z$-scores following $\normal(0,1)$, meaning that we cannot discriminate the samples under detection is whether directly sampled from the distribution of the whole code space $\codeSpace$  or not.

We can notice that this ideal scheme has nearly perfect detection and quality preservation, i.e., $0.9747$ AUROC score and $0.9634$ Pass@1 ($158 / 164$ passed). The left subfigure also shows of \cref{fig:ideal-stat}  that the $z$-score distribution of the ideal watermarked samples is significantly biased from the standard normal distribution, whose significance can be confirmed by $p$-values (right sub-figure of \cref{fig:ideal-stat}), in which 72.78\%  samples have $p$-values lower than 0.02 and 87.97\%  lower than 0.05.

Although the scheme performed ideally, the excellent detection ability is not accompanied by better robustness. After the attack, with a Pass@1 score proven to be maintained at 96.34\%, the 0.5085 AUROC score shows that the detection effect after the attack is like flipping a coin.
Meanwhile, we can notice from \cref{fig:ideal-stat}  that the $z$-scores distributed around zero, and only 7.59\%  (87.97\%  before) cases have $p$-value lower than 0.05 and can be seen as significant.

From the results, we can conclude that even if we exploit the representational power of N-gram features far deeper in the future, the N-gram-based watermarking still has weak robustness confronting our attack.
}

\section{Distribution Consistency Test on the Idealized Watermarking Scheme}
\label{sec:dist_test}
\label{sec:dist_test_app}
{
\newcommand{\normalizer}{\mathsf{norm}}
\newcommand{\deNorm}{\mathsf{deNorm}}
\newcommand{\apprEqSpace}{\eqSpace}
\newcommand{\normCode}{\code_{\mathsf{norm}}}
\newcommand{\expSize}{n}

\begin{algorithm}[!t]
\caption{Equivalent Space}
\label{alg:space_construction}
    \KwIn{
        Seed Code: $\code \in \codeSpace$,  \newline
        Normalizer: $\normalizer: \codeSpace \to \codeSpace$, \newline
        De-normalizer: $\deNorm: \codeSpace \to \codeSpace$, \newline
        Expected Space Size: $\expSize \in \nature$
    } \KwOut {
        Equivalent Space: $\apprEqSpace \subseteq \codeSpace$
    }
    Init Equivalent Space $\apprEqSpace \gets \varnothing$ \\
    Init Normalized Code $\normCode \gets \normalizer(\code)$ \\
    \While{$\sizeof{\apprEqSpace} < \expSize$} { \label{line:check_size}
        Init $\code' \gets \deNorm(\normCode)$ \\
        \If {$\normCode = \normalizer(\code')$} { \label{line:check_norm}
            Set $\apprEqSpace \gets \apprEqSpace \cup \curly{\code'}$ \\
        }
    }
    Return $\apprEqSpace$
\end{algorithm}

\begin{figure}[t]
    \centering
    \includegraphics[width=1\columnwidth, trim={0 2.7mm 0 0}, clip]{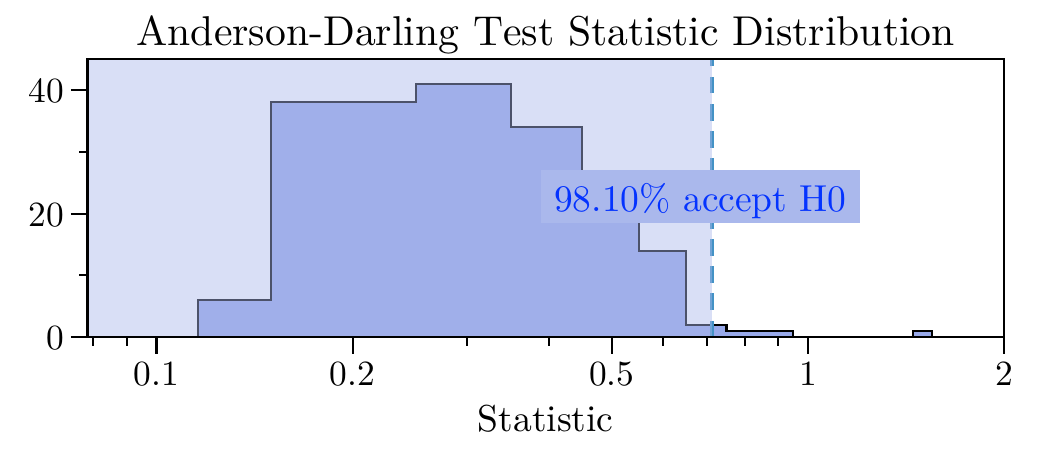}
    \caption{Anderson-Darling statistics distribution for testing whether each approximate equivalent space follows $\normal(0,1)$; the blue line marks significance level $5.0$.}
    \label{fig:anderson-dist}
\end{figure}

\begin{figure}[t]
    \centering
    \includegraphics[width=1\columnwidth, trim={0 2.8mm 0 0}, clip]{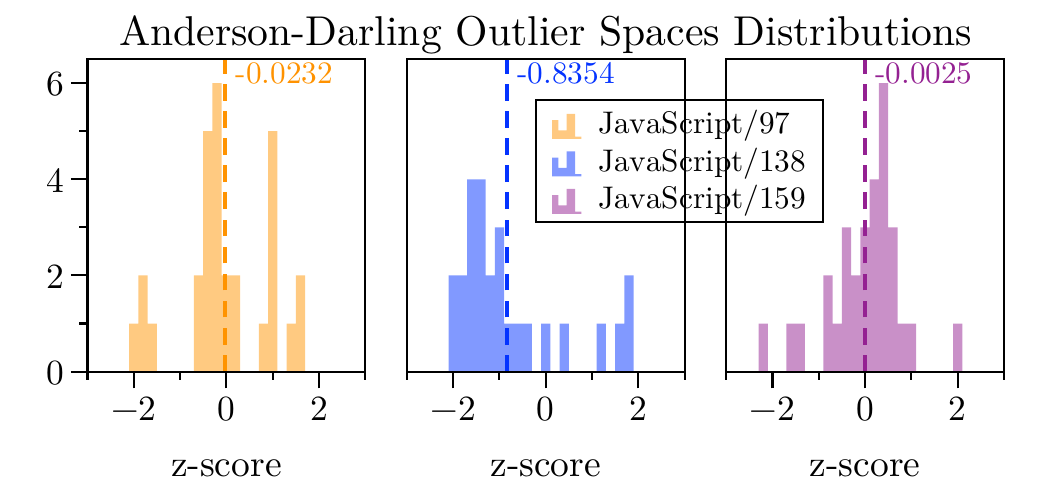}
    \caption{Z-score distributions of the few outlier spaces, with dashed lines marking their means.}
    \label{fig:outlier-spaces}
\end{figure}

We directly test the core assumption behind our theory: distribution consistency (\cref{asm:main}).
Starting from 158 highly watermarked code segments produced by the idealized scheme in \cref{sec:exp_ideal_wm_attack}, we construct approximate rule-based equivalent spaces and test whether their internal $z$-score distributions match the standard normal distribution of the broader code space.

\cref{alg:space_construction} shows the construction of each equivalent space.
Specifically, we use UglifyJS as a code normalizer (we did not find an effective code normalizer) and gpt-4o \cite{hurst2024gpt}  as a randomized de-normalizer.
The algorithm repeatedly de-normalizes the normalized seed code, trying to generate code samples that belong to the same equivalent space as the seed code.
If the de-normalized code has the same normalized form as the seed code (\cref{line:check_norm}), it must belong to the same equivalent space as the seed code, i.e., the normalizer has the ability to transform one into another.
The algorithm will repeat until the size of constructed equivalent space reaches the expected size (\cref{line:check_size}) (30 in practice).
In this way, we use a randomized de-normalizer to select samples from equivalent spaces.

Using the Anderson--Darling test, \cref{fig:anderson-dist} shows that $98.10\%$ of equivalent spaces fail to reject the null hypothesis at the conventional $5.0\%$ significance level.
The remaining three outlier spaces do not indicate signals that the watermark preserved either: one is biased toward lower $z$-scores, and the other two have means close to zero (see \cref{fig:outlier-spaces}).
Even at the more permissive $15.0\%$ significance level~\cite{stephens1974edf}, $83.54\%$ still match $\normal(0,1)$.

In addition, because we use UglifyJS as a normalizer even though obfuscators are not designed for code normalization, our results provide a conservative estimate of distribution consistency compared with using a dedicated normalizer built from various implementable equivalent transformation rules.
These results support that a practical rule set can satisfy distribution consistency to a high degree, leaving little room for the watermark signal to survive within the equivalent space.

}

\section{A Theory-Grounded Random-Walk Obfuscator}
\label{sec:rw_obf}

Our theoretical attack (\cref{tho:main}) is a randomized walk driven by an ergodic rule set, whereas the off-the-shelf tools in \cref{sec:experiment_setting} apply their transformation rules along fixed paths.
To connect the two directly, we implement a Python obfuscator that follows the random-walk construction of \cref{sec:chain_modeling}.

\paragraph{Construction.}
The obfuscator operates with reversible transformation rules, a no-op transition, and a bounded state space, matching the requirements of \cref{def:erg_rule_set}.
For example, variable renaming is made reversible by mapping each identifier's initial name to a finite candidate set and allowing the inverse mapping back to the initial name.
The no-op transition guarantees aperiodicity, and each walk is restricted to a fixed $2{,}000$-character length interval around its seed program, which keeps the reachable state space finite.
The transformation rules are largely adapted from Python-Minifier, extended with two additional sets of reversible rules that implement control-flow flattening and opaque-predicate insertion (i.e., advanced rules).
We run $100$ transformation steps per program; every program that passed its benchmark tests before transformation continued to pass them afterward.

\paragraph{Evaluation.}
We apply the obfuscator to all effective settings from our experiments, defined as those retaining more than $80\%$ of the non-watermarked performance while achieving an AUROC above $0.8$.
This yields $43$ selected Python experimental settings---$13$ for WLLM, $24$ for SWEET, and $6$ for SynthID---with each watermarking configuration counted separately on each dataset, covering $11{,}118$ watermarked code programs in total.
To isolate the contribution of the advanced rules, we also run a variant of the obfuscator with control-flow flattening and opaque predicates disabled, and we report Python-Minifier on the same settings for comparison.

\begin{table}[H]
    \centering
    \footnotesize
    \setlength{\tabcolsep}{3pt}
    \begin{tabular}{l c c c c}
        \specialrule{1.2pt}{0pt}{0pt}
        Obfuscator & WLLM & SWEET & SynthID & Overall \\
        \specialrule{0.9pt}{0pt}{0pt}
        \multicolumn{5}{l}{\textit{Settings with AUROC in $(0.4, 0.6)$}} \\
        Random walk & 13/13 & 24/24 & 6/6 & 43/43 \\
        \quad w/o advanced & 13/13 & 23/24 & 5/6 & 41/43 \\
        Python-Minifier & 13/13 & 24/24 & 6/6 & 43/43 \\
        \hline
        \multicolumn{5}{l}{\textit{Mean AUROC}} \\
        Random walk & 0.513 & 0.507 & 0.531 & 0.512 \\
        \quad w/o advanced & 0.529 & 0.525 & 0.544 & 0.529 \\
        Python-Minifier & 0.506 & 0.505 & 0.519 & 0.507 \\
        \specialrule{1.2pt}{0pt}{0pt}
    \end{tabular}
    \caption{
    Watermark detectability under the random-walk obfuscator, its variant without control-flow flattening and opaque predicates (w/o advanced), and Python-Minifier, over the $43$ effective Python settings ($13$ WLLM, $24$ SWEET, $6$ SynthID).
    The upper block counts settings whose post-obfuscation AUROC falls in the chance-level interval $(0.4, 0.6)$; the lower block reports mean post-obfuscation AUROC.
    }
    \label{tab:rw_obf}

\end{table}

\paragraph{Results.}
As shown in \cref{tab:rw_obf}, the random-walk obfuscator drives all $43$ settings into the chance-level interval $(0.4, 0.6)$, with a mean AUROC of $0.512$.
The theoretically grounded construction therefore breaks the robustness of all three watermarking schemes in practice.
Disabling control-flow flattening and opaque predicates leaves two settings outside the interval, with AUROCs of $0.61$ and $0.64$; the advanced transformations eliminate these remaining outliers and reduce the mean AUROC across all three schemes.
Python-Minifier yields similarly chance-level AUROCs across the same $43$ settings.
Deterministic tools of this kind follow a fixed transformation path through the same rule-based transformation graph modeled by $\ruleSet$; the agreement between the randomized construction and Python-Minifier shows that these fixed paths already realize, in practice, the limiting behavior that \cref{tho:main} establishes for the randomized walk.

\section{Repository-Level Evaluation on SWE-bench Verified}
\label{sec:swebench}

The benchmarks in \cref{sec:experiment_setting} consist of short, single-function programs.
To examine whether our attack remains effective and affordable on longer, repository-level code generation, we evaluate it on SWE-bench Verified \cite{jimenez2024swe, chowdhury2024swebenchverified}.

\paragraph{Setup.}
We randomly sample $50$ tasks from the benchmark and solve them with mini-SWE-agent \cite{yang2024swe, minisweagent2025}, using the official INT4-quantized Qwen3.5-35B-A3B \cite{qwen2026qwen35} as the backbone model.
The non-watermarked agent achieves a $44\%$ solve rate.
We then apply WLLM under each of its $15$ hyperparameter settings (the same grid as in \cref{sec:experiment_setting}), running the same $50$ tasks once per setting, and apply the detector to the lines added by the agent's patch.
Watermark strength trades off sharply against task performance in this agentic setting: the two settings with the highest AUROCs, $0.750$ and $0.774$, achieve solve rates of only $26\%$ and $16\%$, retaining at most $59.1\%$ of the non-watermarked performance.
For the seven quality-preserving settings that retain more than $80\%$ of the non-watermarked solve rate, we use Python-Minifier to obfuscate the methods modified by the coding agent, reconstruct each patch, and re-evaluate the AUROCs.

\begin{table}[H]
    \centering
    \small
    \setlength{\tabcolsep}{5pt}
    \begin{tabular}{l c c}
        \specialrule{1.2pt}{0pt}{0pt}
        Metric & Before obfuscation & After obfuscation \\
        \specialrule{0.9pt}{0pt}{0pt}
        AUROC range & $[0.518, 0.682]$ & $[0.432, 0.556]$ \\
        Mean AUROC & 0.612 & 0.491 \\
        \specialrule{1.2pt}{0pt}{0pt}
    \end{tabular}
    \caption{
    WLLM detectability on SWE-bench Verified before and after obfuscation, over the seven settings that retain more than $80\%$ of the non-watermarked solve rate.
    }
    \label{tab:swebench}

\end{table}

\paragraph{Results.}
\cref{tab:swebench} reports detectability over the seven quality-preserving settings.
Detectability is generally lower on repository-level tasks to begin with, and obfuscation further removes the remaining signal: all seven post-obfuscation AUROCs lie near $0.5$.
Increasing code and task complexity therefore does not restore watermark robustness.
The attack also remains cheap at this scale: obfuscation takes $0.23$ seconds per patch on average using a single CPU core across the seven settings, and under SWE-bench's official resource limits, no previously solved task times out or runs out of memory after obfuscation.

\section{Detection Against Alternative Negative Classes}
\label{sec:neg_class}

In \cref{sec:experiment_setting}, watermark detectability is measured by AUROC between the z-score distribution of the generated samples and the standard normal distribution.
In deployment, a defender may instead calibrate the detector against concrete negative samples, such as non-watermarked model outputs, human-written code, or obfuscated non-watermarked code.
We therefore recompute the AUROC scores for all WLLM Python experiments---the $15$ hyperparameter settings on two models and two Python benchmarks, giving $60$ experiments per negative class---against four such negative classes: non-watermarked LLM outputs; human-written code, taken as the canonical solutions of the benchmarks; and two classes of non-watermarked code obfuscated by each of the two off-the-shelf Python obfuscators, PyMinifier and Python-Minifier.

\begin{table}[H]
    \centering
    \small
    \setlength{\tabcolsep}{1.5pt}
    \begin{tabular}{@{}l c c c@{}}
        \specialrule{1.2pt}{0pt}{0pt}
        Negative class & Mean AUROC & $(.4,.6)$ & $(.45,.55)$ \\
        \specialrule{0.9pt}{0pt}{0pt}
        Non-WM LLM code & $0.800 \rightarrow 0.518$ & 60/60 & 52/60 \\
        Human-written code & $0.805 \rightarrow 0.518$ & 60/60 & 49/60 \\
        + PyMinifier & $0.806 \rightarrow 0.519$ & 60/60 & 53/60 \\
        + Python-Minifier & $0.803 \rightarrow 0.516$ & 60/60 & 52/60 \\
        \hline
        Overall & $0.804 \rightarrow 0.518$ & 240/240 & 206/240 \\
        \specialrule{1.2pt}{0pt}{0pt}
    \end{tabular}%
    \caption{
    AUROC of all WLLM Python experiments recomputed against four negative classes. ``Non-WM'' denotes non-watermarked code; ``+ PyMinifier'' and ``+ Python-Minifier'' denote non-watermarked code obfuscated by the respective tool. Each row aggregates the $60$ WLLM Python experiments ($15$ hyperparameter settings $\times$ $2$ LLMs $\times$ $2$ Python benchmarks); mean AUROC is reported before $\rightarrow$ after obfuscation, and the counting columns refer to post-obfuscation scores.
    }
    \label{tab:neg_class}

\end{table}

Before obfuscation, WLLM remains detectable against every negative class.
For example, with $\delta = 3.0$ and $\gamma = 0.25$, the watermarked model retains $94.90\%$ of the non-watermarked HumanEval Pass@1 while the AUROC is at least $0.937$ for every negative class.
After obfuscation, as \cref{tab:neg_class} shows, the mean AUROC is approximately $0.518$ for every negative class; all $240$ AUROC scores fall within $(0.4, 0.6)$, and $206$ of $240$ fall within $(0.45, 0.55)$.
The loss of watermark robustness under obfuscation is thus consistent across all four negative classes, and our conclusion does not depend on the choice of the negative class.

\section{Evaluation of ACW}
\label{sec:acw}

Beyond the three N-gram-based schemes in our main experiments, we evaluate ACW \cite{guo2025practical}, a recent code watermarking scheme designed for practical and effective detection while preserving functionality.

Under its ACW-s setting, we generate $20$ completions for each of the $164$ HumanEval tasks, $3{,}280$ programs in total, and obfuscate the watermarked outputs with Python-Minifier.
After obfuscation, the detection AUROC drops to $52.92\%$, close to random guessing.
The vulnerability to code obfuscation thus extends to this scheme as well, and we are not aware of current evidence that ACW is robust to code obfuscation.

\end{document}